\documentclass[manuscript]{aastex}

\shorttitle{The  UV and optical \ion{Fe}{2} emission lines in  type 1 AGNs }
\shortauthors{Kova\v cevi\'c-Doj\v cinovi\' c \& Popovi\' c}

\begin{document}


\title{The connections between the UV and optical \ion{Fe}{2} emission lines in type 1 AGNs  }


\author{Jelena 
Kova\v cevi\' c-Doj\v cinovi\'c \altaffilmark{1,2} }
\and 
\author{Luka \v C. Popovi\' c\altaffilmark{1,2} }

\affil{\altaffilmark{1}Astronomical  Observatory,  Volgina  7, 11060  Belgrade, Serbia}
\affil{\altaffilmark{2}Isaac Newton Institute of Chile, Yugoslavia branch}
\email{jkovacevic@aob.bg.ac.rs, lpopovic@aob.bg.ac.rs}


\begin{abstract}

We investigate the spectral properties of the UV ($\lambda\lambda$2650-3050 \AA) and optical ($\lambda\lambda$4000-5500 \AA)
\ion{Fe}{2}  emission features in a  sample of 293 type 1 active galactic nuclei (AGNs) from Sloan Digital Sky Survey (SDSS) database.
We explore different correlations between their emission line properties, as well as the correlations with the other emission lines from the spectral range.
We find several interesting correlations and we can outline the most interesting results as follows. (i) There is a kinematical connection
between the UV and optical \ion{Fe}{2} lines, indicating that
the UV and optical \ion{Fe}{2} lines originate from the outer part of the broad line region,  so-called intermediate line region; (ii) The
unexplained anticorrelations of the optical \ion{Fe}{2} (EW Fe II$_{opt}$) vsersus
EW [\ion{O}{3}] 5007 \AA \ and EW Fe II$_{opt}$ versus FWHM H$\beta$ have not been detected for the UV \ion{Fe}{2} lines; (iii) The significant
averaged redshift in the UV \ion{Fe}{2} lines, which is not present in optical \ion{Fe}{2}, indicates an inflow in the UV \ion{Fe}{2} emitting
clouds, and probably their asymmetric distribution. (iv) Also, we confirm the anticorrelation between the intensity ratio of the optical and UV \ion{Fe}{2} lines and FWHM of H$\beta$, 
and we find the anticorrelations of this ratio with the widths of \ion{Mg}{2} 2800 \AA, optical \ion{Fe}{2} and UV \ion{Fe}{2}.  This indicates a very important role for the column density and
 microturbulence in the emitting gas. 
We discuss the starburst activity in high--density regions of young AGNs as a possible explanation of the detected
optical \ion{Fe}{2}  correlations and intensity line ratios of the UV and optical \ion{Fe}{2} lines.

\end{abstract}



\section{Introduction}

The spectral properties of active galactic nuclei (AGNs) depend on the  physical conditions and geometry
of emitting regions that radiate in a wide  wavelength range.
Among a large diversity of the spectral features in AGN spectra, iron lines 
are one of the most intriguing because there are numerous open questions about their nature \citep[see, e.g.,][]{co2000}.
They can be very intense in the UV and optical part around \ion{Mg}{2} $\lambda$2800 \AA\ and 
H$\beta$.
The mechanism of their excitation,  the explanation of the observed \ion{Fe}{2} strength,
the place of the \ion{Fe}{2} emission region in an AGN structure, and some
correlations observed between \ion{Fe}{2} and some other spectral properties are still a matter of debate \citep[for detailed review see ][]{ko2010}.

It is widely accepted that the iron is mostly produced by the Type Ia supernovae after explosions of long-lived, intermediate-mass binaries. 
Therefore, it is expected that the ratio of Fe to some elements (O, N, Mg) that are produced in 
 explosions of 
short-lived, massive stars (primarily Type II supernovae), 
could be a cosmological metallicity indicator, due to their different enrichment timescales \citep{ve2003}; that is
the iron lines may serve to constrain the 
age of an AGN and its host galaxy \citep[see e.g.][and references therein]{do2011, dr2011}.

To understand the nature and 
evolution of AGNs, efforts have been made to investigate the correlations between the spectral properties
in different AGN spectral bands and to determine the physics that is behind the detected correlations 
\citep[see e.g.][etc.]{bg92,w99,cr02,sh03,y04,g04,w06,wa09,lu09,ko2010,pk11,ma12,gr2015}. Some correlations between
 the optical \ion{Fe}{2} and other spectral properties in AGNs have been reported, but the physical 
explanation is unknown \citep[see][]{bg92, ko2010}. 
Some of these correlations, for example, are part of Eigenvector 1 of \citet{bg92}   \citep[for a review see][and references therein]{ko2010, pk11}. 
Among them, the most interesting are the anticorrelations of the equivalent widths (EW) \ion{Fe}{2} optical lines with the EW [O III] and H$\beta$ width. 
 However, it is difficult to determine the physics that is behind these correlations because choosing
 a sample of AGNs using different spectral criteria (continuum luminosity, 
[\ion{O}{3}] strength, FWHM H$\beta$ etc.) can give different correlations between spectral properties, and in some cases even the
opposite in different subsamples \citep{y04,g04,lu09,su09,pk11}. 
As an example, a significant difference is seen between the correlations in 
spectral properties for objects divided by FWHM of H$\beta$  \citep{su09,ko2010}. However, it seems that more
relevant is to consider AGNs with  different [\ion{O}{3}] 5007 \AA \ to narrow H$\beta$ ratios ($\mathrm{[O III]}/\mathrm{H\beta}_{NLR}$), because this may  give an additional information about
the starburst (SB) fraction in AGNs \citep[see][]{pk11}, which is probably related to the AGN evolution \citep{lt06,m09,sa10,pk11}. 
In an early phase of their evolution, AGNs are probably composite objects, which consist of SBs 
(star-forming regions) and the central AGN engine. The 
influence of SBs on the spectral properties becomes weaker in a later phase of the AGN evolution \citep{m09}.

The origin of the iron lines is also very intriguing question. The mechanism of their excitation, 
the explanation of the observed \ion{Fe}{2} strength, and 
the place of \ion{Fe}{2} emission region in an AGN structure are still a matter of debate. 

Several authors have shown that a classical photoionization model cannot sufficiently explain the observed UV and optical \ion{Fe}{2} emission strengths, and 
that additional processes must be included \citep[][etc.]{co1980,jo1987,sp1998,sp2003,co2000,ve2003,ba04,bv2008,sa2011}. There are 
some indications that microturbulence may have a significant influence on the \ion{Fe}{2} strength 
\citep{ne1983,ve2003,sp2003,ba04,bv2008,sa2011}. \citet{ba04} showed that a photoionization model may reproduce well 
the observed shape and the EW of the UV \ion{Fe}{2} 2200-2800 bump, but only if a microturbulent gas motion is taken into account.
The \ion{Fe}{2} strength is also controlled by the column density as well \citep{jo1987,  ve2004, fe2009, sa2011} and that 
the strong \ion{Fe}{2} emission is connected with high density emitting regions 
\citep{jo1991, ba1996, law1997, ku2000, gr1996, ha2013, cl2013}.

It seems that there are significant differences in the physics of the UV \ion{Fe}{2} and optical  \ion{Fe}{2} emission region:
the optical and UV \ion{Fe}{2} lines correlate differently 
with some physical properties. The $\mathrm{Fe II}_{opt}/\mathrm{Fe II}_{UV}$ ratio depends on column density \citep{jo1987, sa2011} and microturbulence \citep{ve2003}.
Classical photoionization models, assuming a 
 symmetric distribution of emitters, fail to account for this ratio. They cannot explain the
 larger-than-predicted ratios of $\mathrm{Fe II}_{opt}/\mathrm{Fe II}_{UV}$ emission. 
 \citet{sa2011} suggested that this failure 
may be caused by some alternative heating mechanisms for the optical \ion{Fe}{2}, as for example, heating by shocks
or a wrong assumption that the  \ion{Fe}{2} 
emission is isotropic \citep[see][]{fe2009}. If the \ion{Fe}{2} clouds are distributed asymmetrically, the observed 
$\mathrm{Fe II}_{opt}/\mathrm{Fe II}_{UV}$  ratio may be 
reproduced \citep{fe2009}.
Moreover,  \cite{fe2009} showed that the UV \ion{Fe}{2} emission is emitted less isotropically than  the optical \ion{Fe}{2}
lines, and that the predicted emission  ratios from the shielded face are  in a good agreement with observations. This asymmetrical distribution is based on the fact that the 
 optical \ion{Fe}{2} lines are, on average, slightly
redshifted, indicating an inflow in the emitting region \citep{hu2008b}. However, \cite{su12} demonstrated that this 
redshift is not significant and  should be taken with caution.

In several previous papers \citep[see][]{po2009,ko2010,pk11,sh2012,po2013}, we investigated the optical \ion{Fe}{2} lines and 
their spectral 
properties in AGNs. Here, we extend our investigation to the UV part of AGN spectra.
The aim of this work is to investigate  relationships between the optical and UV \ion{Fe}{2} emission in order to 
understand the  physics of their emission regions. 
For this purpose, we model the \ion{Fe}{2} lines in the UV and optical bands, and fit the observed spectra with the model. After that,
we explore the correlations between the spectral properties of the optical and UV \ion{Fe}{2} lines, and the correlations between them and the \ion{Mg}{2} and H$\beta$ lines. 
 The flux ratios of the considered lines, which may be indicators of the some physical conditions, are analyzed, as well as the possible connection of 
starburst activity with some unexplained correlations of the iron lines.
 
 The paper is organized as follows: In Section \S 2 we describe the sample selection, spectra decomposition, and method of analysis.
 The results of the performed correlations  and our analysis are given in Section \S 3, and 
discussed in Section \S 4. Finally, in Section \S 5, we outline our conclusions.

\section{The sample and analysis}

\subsection{The  AGN sample}

For this investigation we use the spectra from the Sloan Digital Sky Survey (SDSS), Data Release 7 \citep[see][]{ab2009}. 
The SDSS uses the 2.5 m telescope at the Apache Point Observatory, a pair of spectrographs fed by optical fibers and 120-megapixel 
CCD camera. Data Release 7 (DR7) is the seventh major data release and provides $\sim$ 120,000 QSO spectra.

In order to investigate the correlations between the properties of the \ion{Fe}{2} emission lines in the UV and the optical band of AGN spectra,
we chose the sample with the appropriate redshift range to cover the lines  that are appropriate for this research.  We focus on the following \ion{Fe}{2} multiplets: 27, 28, 37, 38, 41, 42, 43,
48, and 49 (in the optical band, near H$\beta$) and 60, 61, 62, and 63 (in the UV band, near \ion{Mg}{2} 2800 \AA).  The lines that overlap with UV/optical iron lines, \ion{Mg}{2} and H$\beta$, have complex shapes, i.e. they 
consist of several line components that arise in different emission line regions. Therefore, their components are used for comparison with the kinematic and physical properties of the 
\ion{Fe}{2} UV/optical lines, in order to investigate the \ion{Fe}{2} emission region. 

To obtain the sample of AGN spectra from the SDSS database, we use SQL (Structural Query Language) search. The final sample of spectra is chosen using the following criteria.
\begin{enumerate}
\item  A Type 1 AGNs (i.e. a broad line AGNs), classified as a QSO in the  SDSS spectral classification.
 \item A relatively high signal to noise ratio ($\mathrm{S/N}>25$).
\item A good pixel quality.
 \item The redshift within the $0.407\leq z\leq 0.643$ range in order to cover both the optical \ion{Fe}{2} lines around H$\beta$ and 
 \ion{Fe}{2} UV lines around \ion{Mg}{2} 2800 \AA.
\item A high redshift confidence (zConf$>$0.95).
\item The presence of the broad H$\beta$ and \ion{Mg}{2} 2800 \AA \ (their equivalent widths should be larger than zero).
\item There is no any absorption in the \ion{Mg}{2} and UV \ion{Fe}{2} lines.
\end{enumerate}
Our sample contains 293 AGN spectra, which are used for this investigation.

The correction for Galactic extinction is made by using the standard Galactic-type law (\citealt{se1979} for the
UV, \citealt{ho1983} for optical-IR) and Galactic extinction
coefficients given by \citet{sc1998}, which are available from the NASA/IPAC Extragalactic 
Database\footnote{http://nedwww.ipac.caltech.edu/}. 

\begin{figure*}

\includegraphics[width=0.50\textwidth]{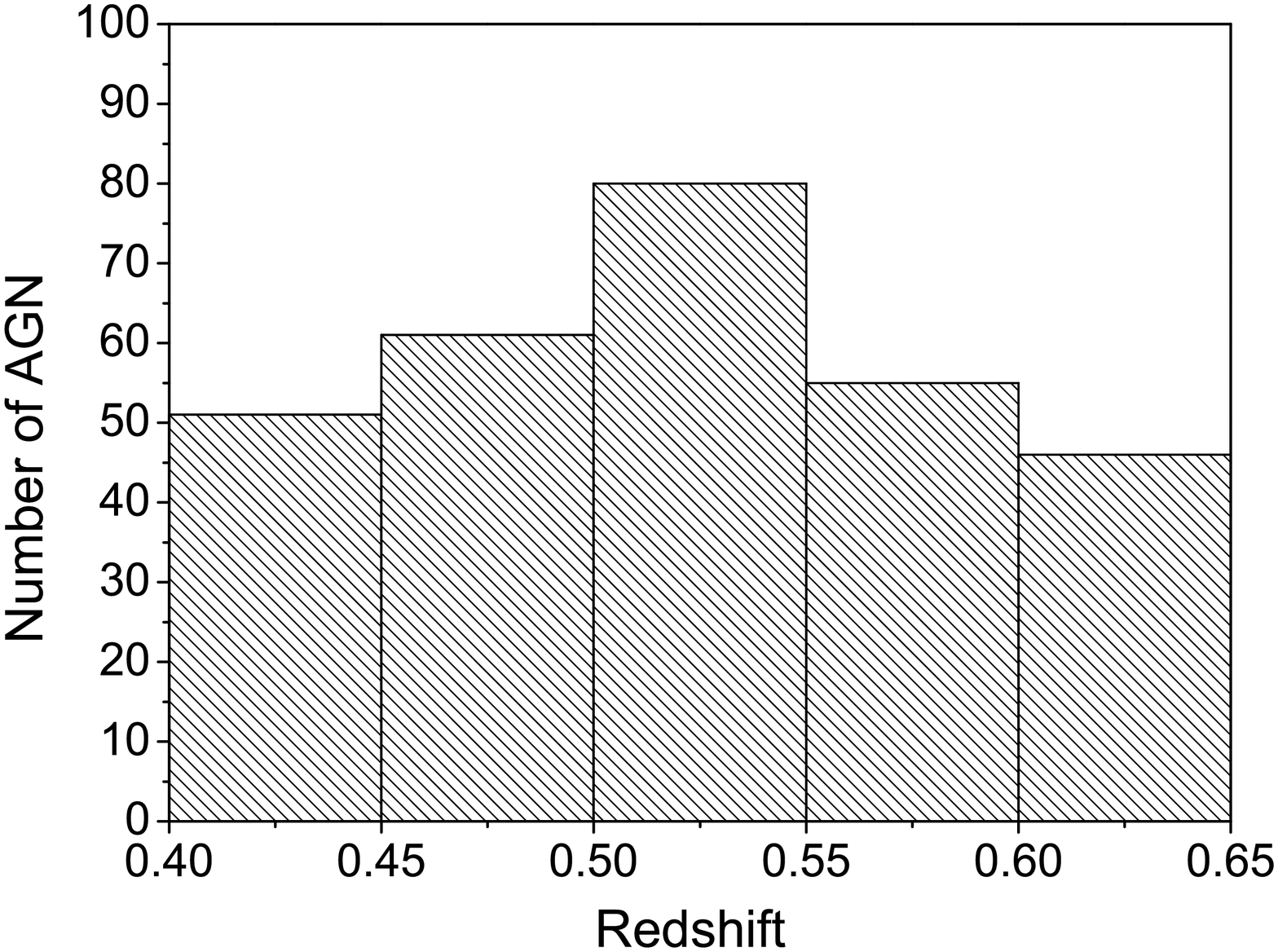}
\includegraphics[width=0.50\textwidth]{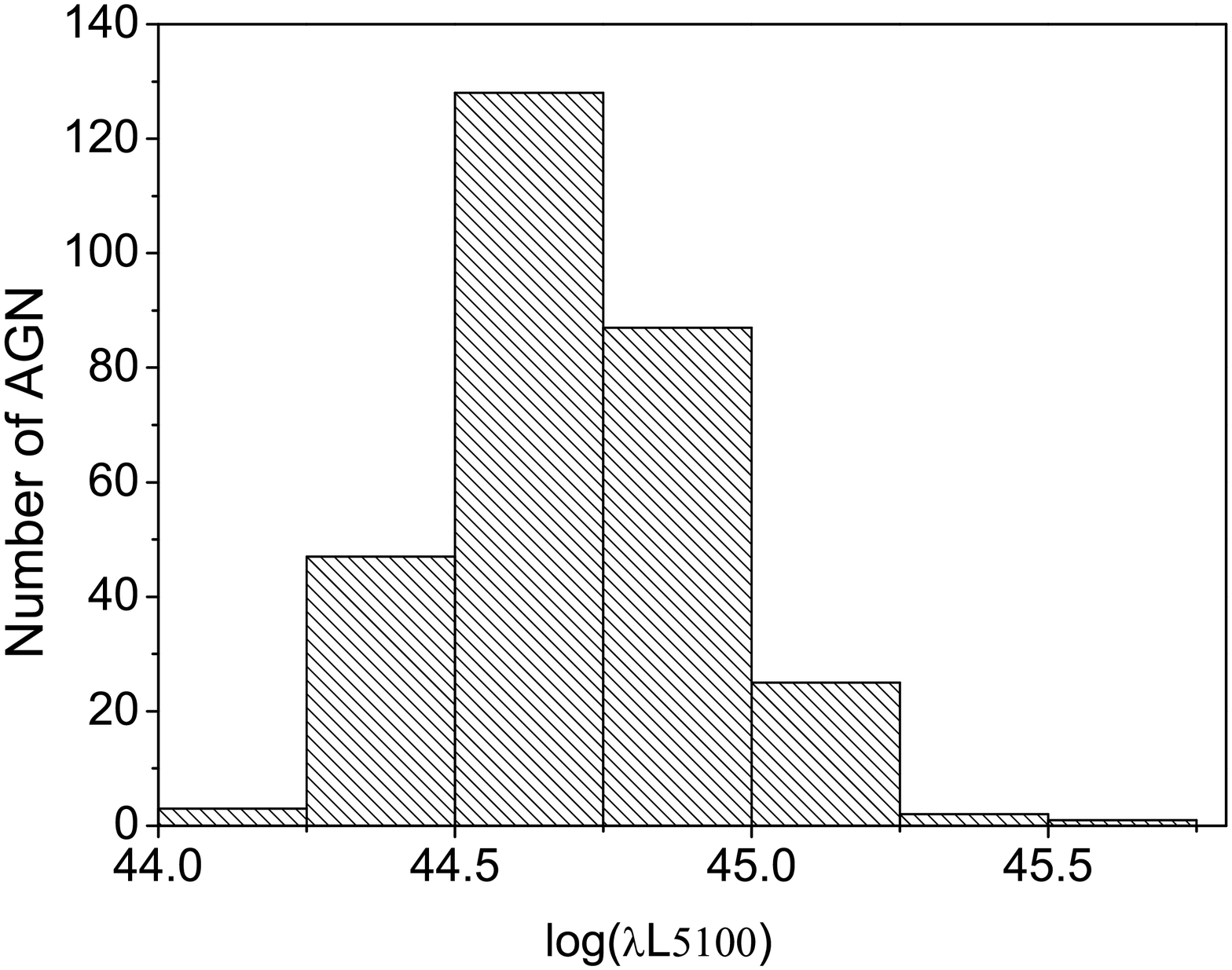}
\caption{Distribution of the redshift (left) and luminosity (right) in the sample of 293 AGNs.}
\label{0}
\end{figure*}

The luminosity and redshift distributions for the final sample are given in Fig \ref{0}. As it can be seen, the redshift 
distribution is approximately uniform (between 0.4 and 0.65), but majority of the objects ($\sim$ 75 \%) have luminosity in 
a very narrow range: 44.5$<$log($\lambda$L$_{5100}$)$<$45. Luminosities were calculated using the formula given in \citet{b306}, 
with adopted cosmological parameters of  $\Omega_M$ = 0.27, $\Omega_\Lambda$ = 0.73, $\Omega_k$ = 0, and Hubble constant $\mathrm{H_o=71\ \rm kms^{-1}Mpc^{-1}}$.
The luminosity of the continuum is taken to be an average value in the interval of $\lambda\lambda$ 5100-5105 \AA.

  We also consider the $\mathrm{[O III]_{5007}/H\beta_{NLR}}$ ratio, since it may  give some indication of
the SB activity in the central part of AGN \citep[see][]{pk11}
assuming that $\mathrm{log([O III]/H\beta}_{NLR}\mathrm{)<0.5}$  indicates SB-dominant objects, and 
the $\mathrm{log([O III]/H\beta}_{NLR}\mathrm{)>0.5}$ indicates AGN-dominant objects. We found only 46
objects with $\mathrm{log([O III]/H\beta}_{NLR}\mathrm{)<0.5}$, which is not statistically significant in the sample. Therefore,
we did not perform correlations for each  subsample, but only plot them with different notations in figures where difference between their properties is easily shown.
 In this way, we try to see whether AGN evolution (which is probably related with the presence/absence of the
starburst regions) has any influence on these spectral correlations.

\subsection{The broad line and continuum model \break in the spectral range $\lambda\lambda$ 2650-5500 \AA}

We  use a model that consists of the UV-optical continuum, the \ion{Fe}{2} templates, and complex shapes of  broad lines.
The model is applied within a wide spectral range from 2650 \AA\ to 5500 \AA. 
We explore  the UV \ion{Fe}{2} lines ($\lambda\lambda$ 2650-3050 \AA), which are in the \ion{Mg}{2} 2800 \AA \ spectral range, and the optical \ion{Fe}{2} lines ($\lambda\lambda$ 4000-5500 \AA)
which are in the Balmer lines (H$\beta$, H$\gamma$, H$\delta$) spectral range.

\subsubsection{The line and continuum model in the optical ($\lambda\lambda$ 4000-5500 \AA) range}

To fit the optical  emission lines, the optical continuum is estimated using the continuum windows given in 
 \citet{ku2002}. The points of the continuum level are interpolated and the continuum is subtracted. 
 After that, emission lines in the  $\lambda\lambda$ 4000-5500 \AA \ range 
are fitted with a model of multi-Gaussian functions \citep{po2004}, where each Gaussian is assumed to represent emission from one emission region. 
The width and shift of each Gaussian reflects the kinematical properties of an emission region 
 \citep[see][and references therein]{ko2010}. 

All narrow lines from the spectra are assumed to have the same velocity dispersion and velocity shift, because it is assumed that
they are
originating  in the same emission region, the Narrow Line Region (NLR).
Consequently, parameters of the widths and shifts of the narrow lines are taken to be the same for the 
[\ion{O}{3}] $\lambda\lambda$4959, 5007 \AA, [\ion{O}{3}] $\lambda$4363 \AA  \ lines, as well as for the narrow 
components of the Balmer lines. The [\ion{O}{3}] $\lambda\lambda$4959, 5007 \AA \ lines are fit with an additional
component that describes the asymmetry  in the wings of these lines \citep[see][]{ko2010}. The ratio of the 
[\ion{O}{3}] $\lambda\lambda$4959, 5007 \AA \ has been taken as 1:3 \citep[see][]{dim07}.

\begin{figure*}

\includegraphics[width=0.60\textwidth,angle=270]{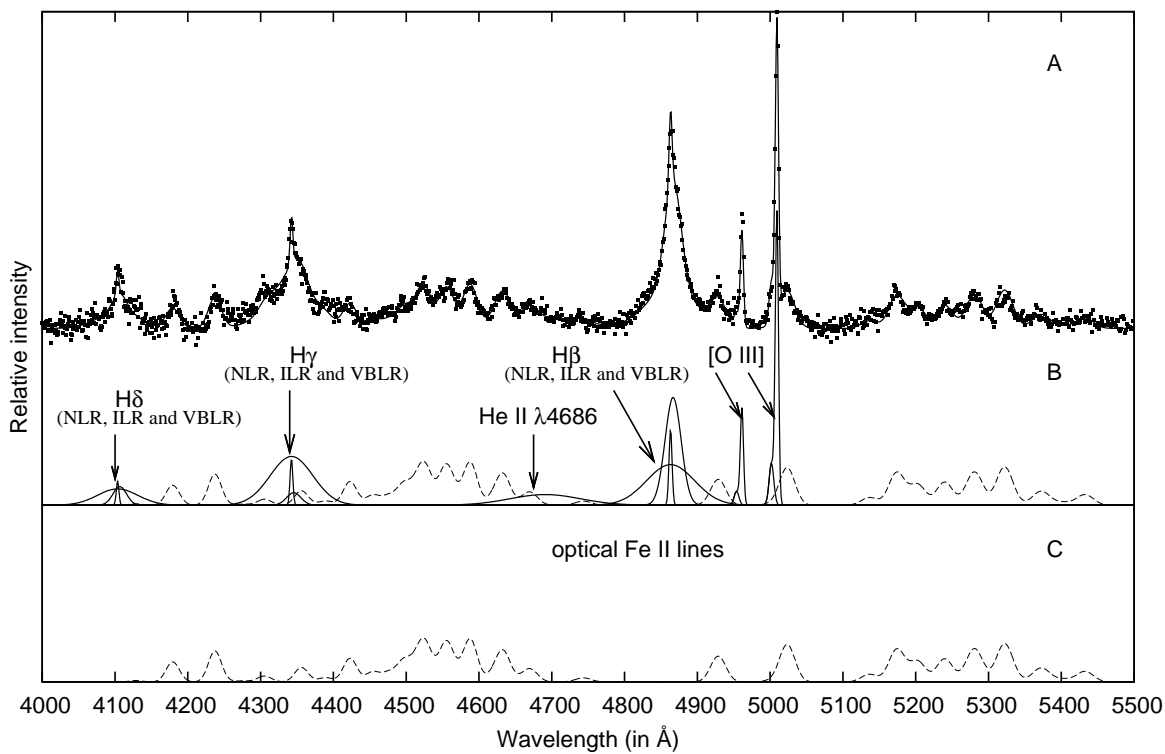}
\caption{Example of the fit of spectrum of SDSS J020039.15$-$084554.9 in
the $\lambda\lambda$ 4000-5500 \AA \ region. A: The observed spectrum (dots) and the best
fit (solid line). B: Decomposition of the emission lines. \ion{Fe}{2} template is denoted
with dashed line. C: The \ion{Fe}{2} template is shown separately.}
\label{1}
\end{figure*}

The  H$\beta$ line is fit with three Gaussians: 
one represents the emission from the NLR, and other
two represent  the emission from the Broad Line Region (BLR)--
the Gaussian that fits the line core of H$\beta$ is assumed to be 
emission from the outer part of the BLR  (Intermediate Line Region - ILR) and one 
that fits the line wings is assumed to be emission coming from the deeper layers of the BLR, closer to the black 
hole (Very Broad Line Region - VBLR). 
Therefore, the H$\beta$ line is decomposed into three Gaussian components: NLR, ILR, and VBLR
\citep[][etc.]{br1994, co1996, po2004, il2006, bo2006, bo2009, hu2008b, ko2010, zh2011, hu2012}.
The H$\gamma$ and H$\delta$ lines are fitted in the same way as  H$\beta$, assuming that their components have 
the same widths and shifts as the corresponding components of H$\beta$ (see Fig. \ref{1}). 
The intensities of the NLR, ILR, and VBLR components are taken to be the free parameters for all Balmer lines. 

 The \ion{He}{2} $\lambda$4686 \AA \ line is fitted with one broad Gaussian. The numerous optical iron lines in the 
 $\lambda\lambda$ 4000-5500 \AA \ range are fitted with template given by \citet{ko2010}, and extended for the 
 \ion{Fe}{2} lines near $\sim\lambda$4200 \AA \ \citep{po2013, sh2012}\footnote{The \ion{Fe}{2} template  
 $\lambda\lambda$ 4000-5500 \AA, as well as the web application for fitting on-line \ion{Fe}{2} lines with 
 this model are given at http://servo.aob.rs/FeII\_AGN/ as a part of Serbian Virtual Observatory.}. 
 The ${\chi}^2$ minimization routine is applied to obtain the best fit \citep{po2004}. An example of 
 the best fit in the optical part of spectra is shown in Fig. \ref{1}.

\subsubsection{The UV Balmer pseudocontinuum}

In order to fit the lines in the UV range ($\lambda\lambda$ 2650-3050 \AA), first one needs to model the UV Balmer 
pseudocontinuum. 
The UV Balmer pseudocontinuum consists of the power law and the bump at 3000 \AA, which represents the sum of the blended, broad, high-order 
Balmer lines and the Balmer continuum. We fit simultaneously the power law 
and the Balmer continuum (together with high order Balmer lines), using the model described in \citet{ko2014}. 
This model consists of the function given in \citet{gr1982} for the Balmer continuum, in the case of a partially optically 
thick cloud, but with one degree of freedom less: for the intensity of the Balmer continuum, which is calculated using 
the prominent Balmer lines in the spectra.  In this way, the less uncertain estimation of the Balmer continuum is achieved. 

The Balmer continuum intensity at the Balmer edge ($\lambda$ = 3646 \AA) is equal to the sum of the intensities of all 
high order Balmer lines
at the same wavelength ($\lambda$ =  3646 \AA). The broad component of each Balmer line is roughly described with only one
Gaussian, which has the same width and shift for all Balmer 
lines, and their relative intensities are taken from the literature or calculated \citep[see][]{ko2014}. 
Then, if the width, shift, and intensity of only one broad Balmer line (e.g. H$\beta$) are obtained from the fit and the sum of
the fluxes of all high-order Balmer lines that contribute to the Balmer edge have been calculated, then we can obtain 
the intensity of the Balmer continuum at the Balmer edge.
The model is applied for the uniform temperature T$_{e}$=15 000 K and
  optical depth at the Balmer edge fixed at: $\tau_{BC}$=1  \citep[see][]{ku2007}. Using this model, the pseudocontinuum 
  is fitted with four free parameters: the width, shift and intensity of the one of prominent Balmer line (in our case H$\beta$
  or H$\gamma$), and the exponent of the
  power law.

\begin{figure*}

\includegraphics[width=0.75\textwidth]{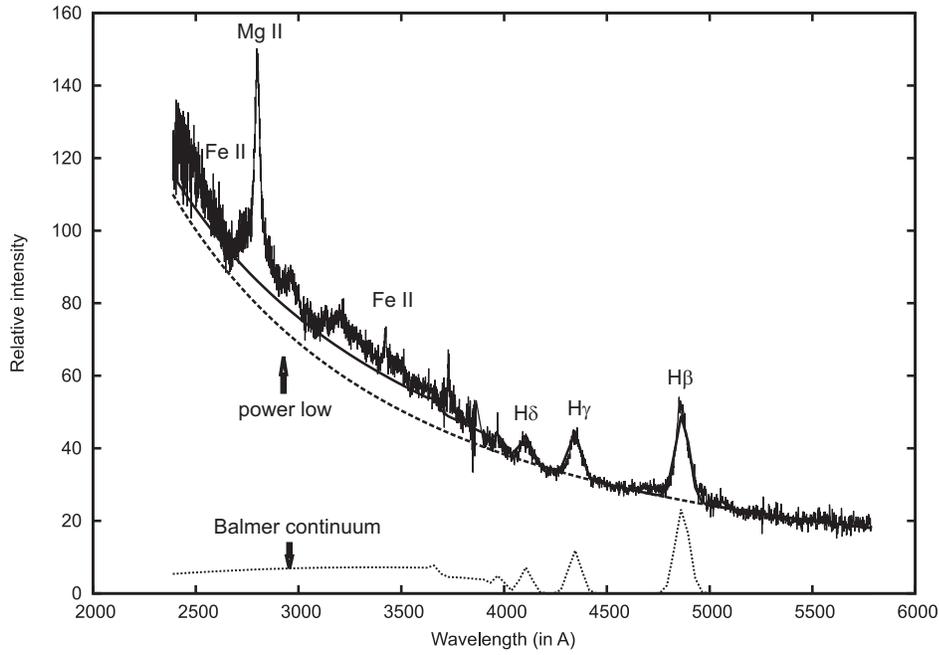}
\caption{Example of the UV continuum fit (SDSS J092423.42$+$064250.6) with the power law and
Balmer continuum model: dotted line - 
Balmer continuum with high order Balmer lines, dashed lines - power law and solid line - sum of Balmer continuum, 
high order Balmer lines and power law. The \ion{Fe}{2} optical lines, [\ion{O}{3}] lines and narrow components of Balmer lines 
are removed from the spectrum.}
\label{2}
\end{figure*}

To apply the Balmer continuum model, it is important to have a clean profile of strong broad Balmer lines without any 
contamination from lines that overlap with them (optical \ion{Fe}{2} and [O III]), as well as without the narrow component of 
Balmer lines. We used the fitted data in the optical range to subtract the narrow components and satellite lines, and to obtain the 
 broad Balmer line profile. Then, we applied the same procedure for fitting the UV-pseudocontinuum as 
described in \citet{ko2014}. An example of the UV-pseudocontinuum fit is shown in Fig \ref{2}.

 Because it is very important to correctly subtract the Balmer
continuum in order to measure well the equivalent widths (EWs) of the UV lines, the applicability of the model is tested using the total sample of 293 AGNs chosen for this investigation. 
We measured the dif\-ference between the observed f\-lux and calculated f\-lux (Balmer continuum + power law), and  results are presented in \citet{ko2014}, Section 3.
 We found that the discrepancy between the observed and calculated flux in the UV (at $\sim$2650 \AA) is smaller than 10\% for 92\%  of the sample. This means that for the majority 
of the objects from the sample, the EWs are probably measured well. The other 8\% of objects, with uncertain continuum determination, do not affect the final result.

\subsubsection{The model of the line spectra in the UV ($\lambda\lambda$ 2650-3050 \AA) range}

After determination and subtraction of the UV-pseudocontinuum, the \ion{Mg}{2} 2800 \AA \ line and \ion{Fe}{2} template are simultaneously fit. 

Note that the \ion{Mg}{2} 2800 \AA \ line is the resonant doublet \ion{Mg}{2} $\lambda\lambda$ 2795, 2803 \AA,  where two lines
of the doublet cannot be resolved because of their very large widths. 
The doublet is observed in the spectra of the Type 1 AGNs as a broad,
single line. It is very difficult to find an appropriate model to fit the components of the doublet 
because their relative intensities are
not fixed, i.e. their flux ratio depends on optical depth in the line. In general, one can expect a doublet ratio from
optically thick gas to be approximately in the range $\lambda$2795/$\lambda$2803$\approx$2:1 to 1:1 \citep{la1997}. 
Additionally, it is not possible to get an unique 
Gaussian decomposition because two 
broad \ion{Mg}{2} doublet components overlap with central wavelength difference
($\sim\Delta\lambda\approx$ 8 \AA).
In order to make the fitting procedure more simple, we fit the \ion{Mg}{2} doublet as a single \ion{Mg}{2} 2800 \AA \ line with
two Gaussians: one that fits the core and one that fits the wings of the \ion{Mg}{2} 2800 \AA \  line. In this way, 
Doppler widths of the 
Gaussians that fit the core and wings of the single \ion{Mg}{2} 2800 \AA \  line overestimate the Doppler widths of 
the \ion{Mg}{2} doublet components
for $\approx$ 260 km s$^{-1}$ ($\sim$8 \AA \ separation). Because the widths of the \ion{Mg}{2} lines are generally one 
order of magnitude larger 
than this value, we assume that it is in the range of the error-bars.

\subsubsection{The UV \ion{Fe}{2} line emission model}

The numerous UV \ion{Fe}{2} lines are fitted with the model described in \citet{po2003}. In this model, 
the strongest UV \ion{Fe}{2} lines, within $\lambda\lambda$ 2650-3050 \AA \ 
range, are divided into 4 multiplets: 60 ($\lambda\lambda$ 2907-2979 \AA), 61 ($\lambda\lambda$ 2861-2917 \AA), 
and additionally with 62 and 63 which overlap at $\lambda\lambda$ 2709-2749 \AA. 
 The lines are fitted with 4 parameters of the intensity, for each multiplet. Within one multiplet group, 
 the relative intensities of the lines are fixed using the line strength from NIST\footnote{http://www.nist.gov/pml/data/asd.cfm}
 \citep[see][]{po2003}. It has been assumed that 
 all UV \ion{Fe}{2} lines in this range are originating in the same emission region and consequently
  to have the same Doppler width and shift.
 Therefore, the UV \ion{Fe}{2} template consists of 6 free parameters in the fitting procedure (4 parameters of 
 intensity, width and shift).

The list of the lines and multiplets of the UV \ion{Fe}{2} template, as well as the transitions and relative intensities 
within each multiplet, are given in Table \ref{tbl-1}.  The most 
intensive line within each multiplet is scaled to the unit intensity.

\begin{figure*}

\includegraphics[width=0.60\textwidth]{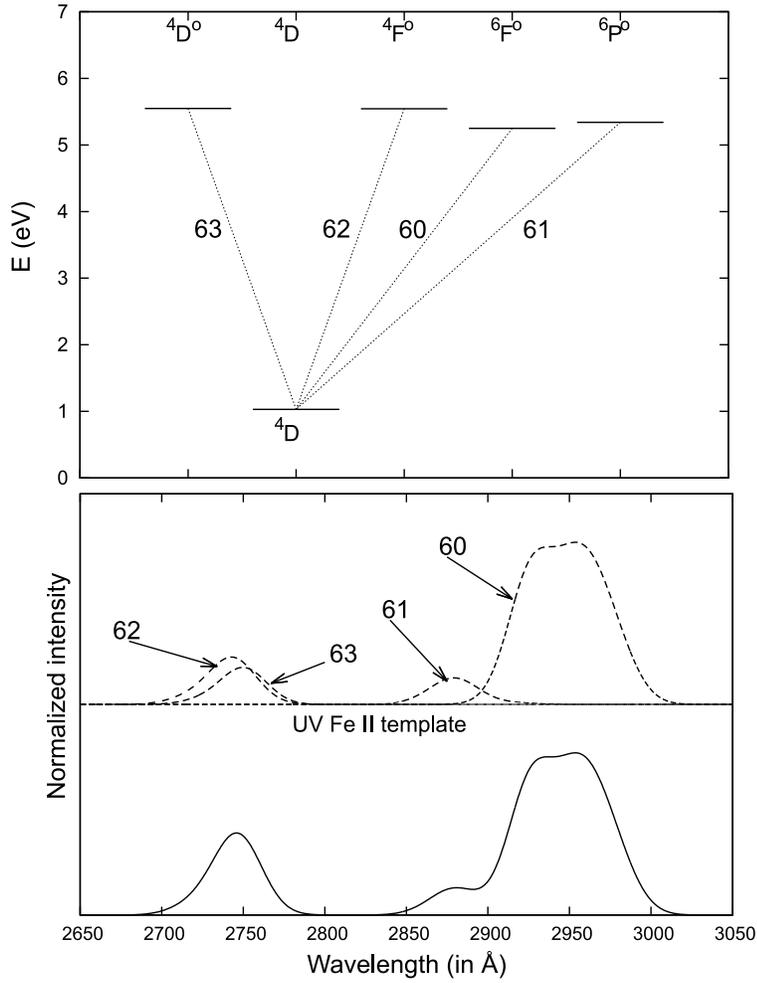}
\caption{Strongest \ion{Fe}{2} multiplets within the $\lambda\lambda$ 2650-3050 \AA \ wavelength range are shown in the 
Grotrian diagram (top) and in the spectrum (middle). The total UV \ion{Fe}{2} template is shown in the bottom. 
In this example, the intensity ratio for multiplets 60, 61, 62, 63 is taken to be: 10:5:0.03:0.07, respectively.}
\label{3}
\end{figure*}

\begin{figure*}
\includegraphics[width=0.47\textwidth]{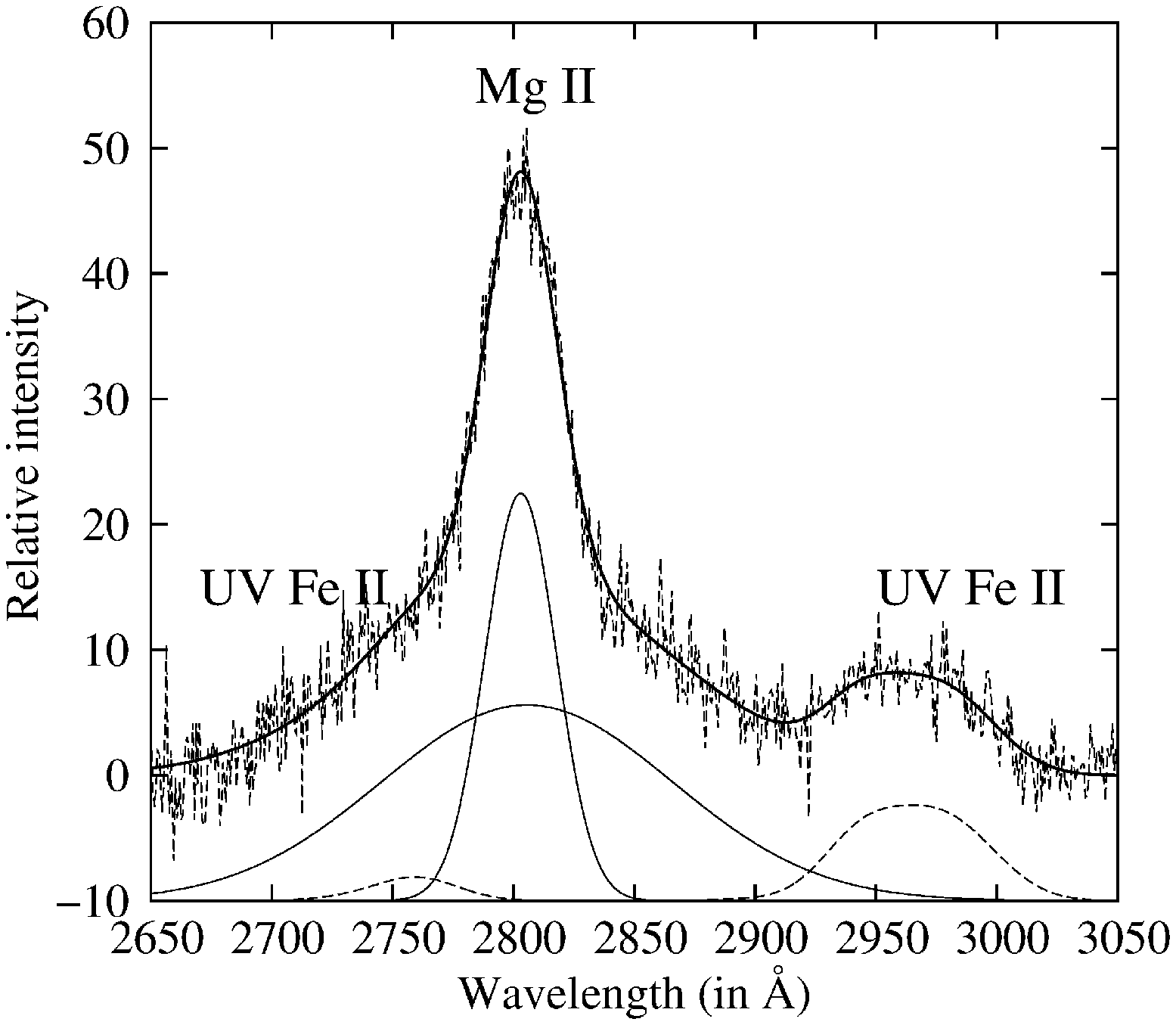}
\includegraphics[width=0.47\textwidth]{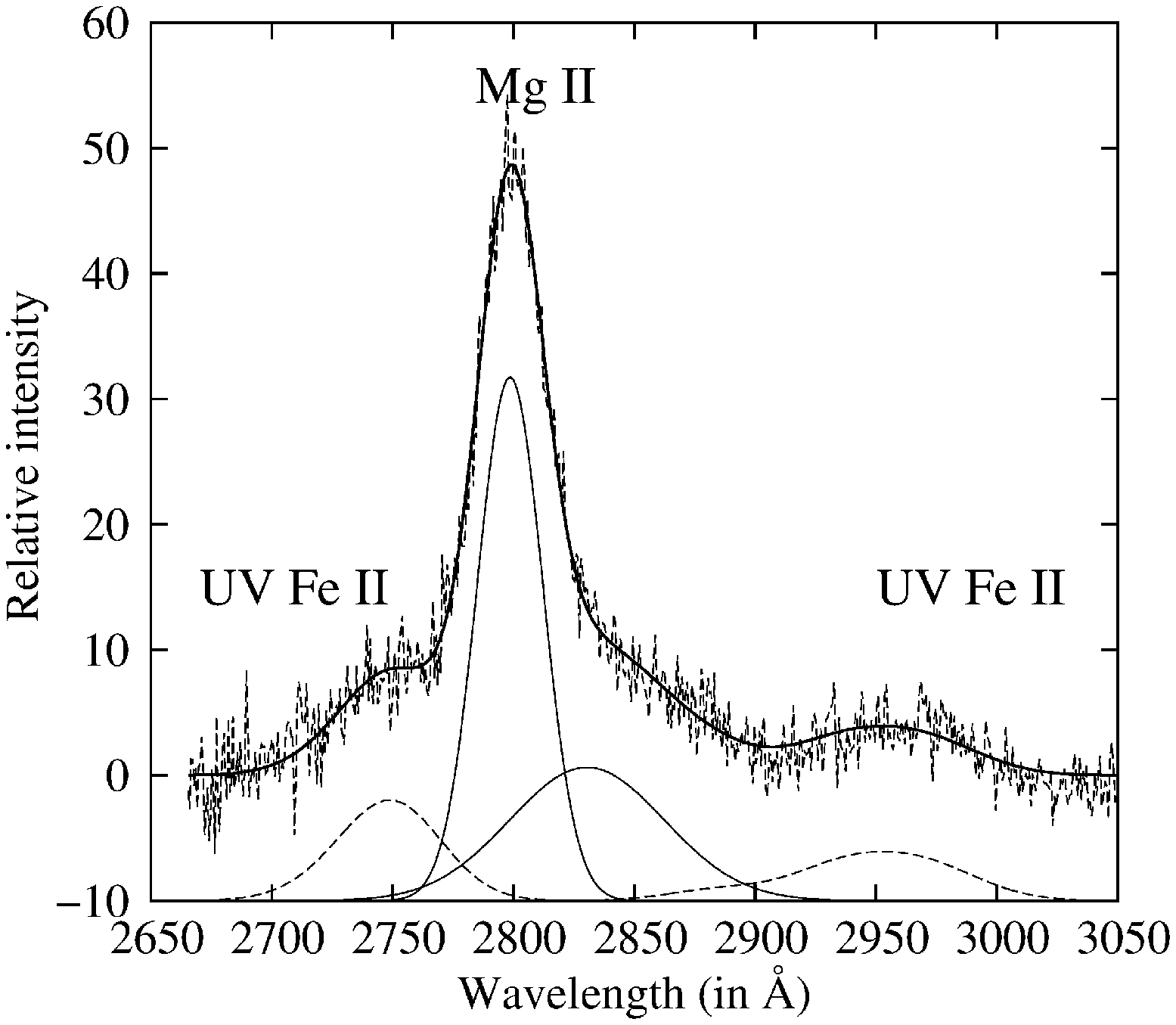}
\caption{Examples of the fit of \ion{Mg}{2} and \ion{Fe}{2} lines in the UV range. The best fit is denoted with solid line. The \ion{Mg}{2} 2800 line is 
fitted with two Gaussians: one fitting the core and one the wings. 
The UV \ion{Fe}{2} template is denoted with the dashed line. The flux of the \ion{Fe}{2} multiplets 60 and 61 are stronger 
compared with 62 and 63 for SDSS J095758.44-002354.0 (left) and it is opposite for the 
object SDSS J155534.61$+$345948.9 (right). }
\label{4}
\end{figure*}

In Fig \ref{3}, the multiplet transitions that are included in the template are presented as the Grotrian diagram
and shown as a spectrum. The examples  of the fitted emission lines (\ion{Mg}{2} and UV Fe II)   in the  UV range 
 are shown in Fig \ref{4}. As it can be seen in Fig. \ref{4}, there is a significant 
difference between the intensity of UV Fe II
multiplets for different AGN spectra.

\subsection{The line parameters}

In order to investigate the spectral properties, we obtain the equivalent widths (EWs) of all considered lines and their
components. EWs have been measured with respect to the continuum below the lines, after subtraction of all satellite
lines. In the case of the UV lines, the intensity of Balmer continuum is estimated first. After that, the Balmer continuum is subtracted
and the EWs of the UV lines (Mg II, \ion{Fe}{2} UV) are measured with respect to the power law continuum component below the lines. 

\begin{figure*}
\includegraphics[width=0.45\textwidth]{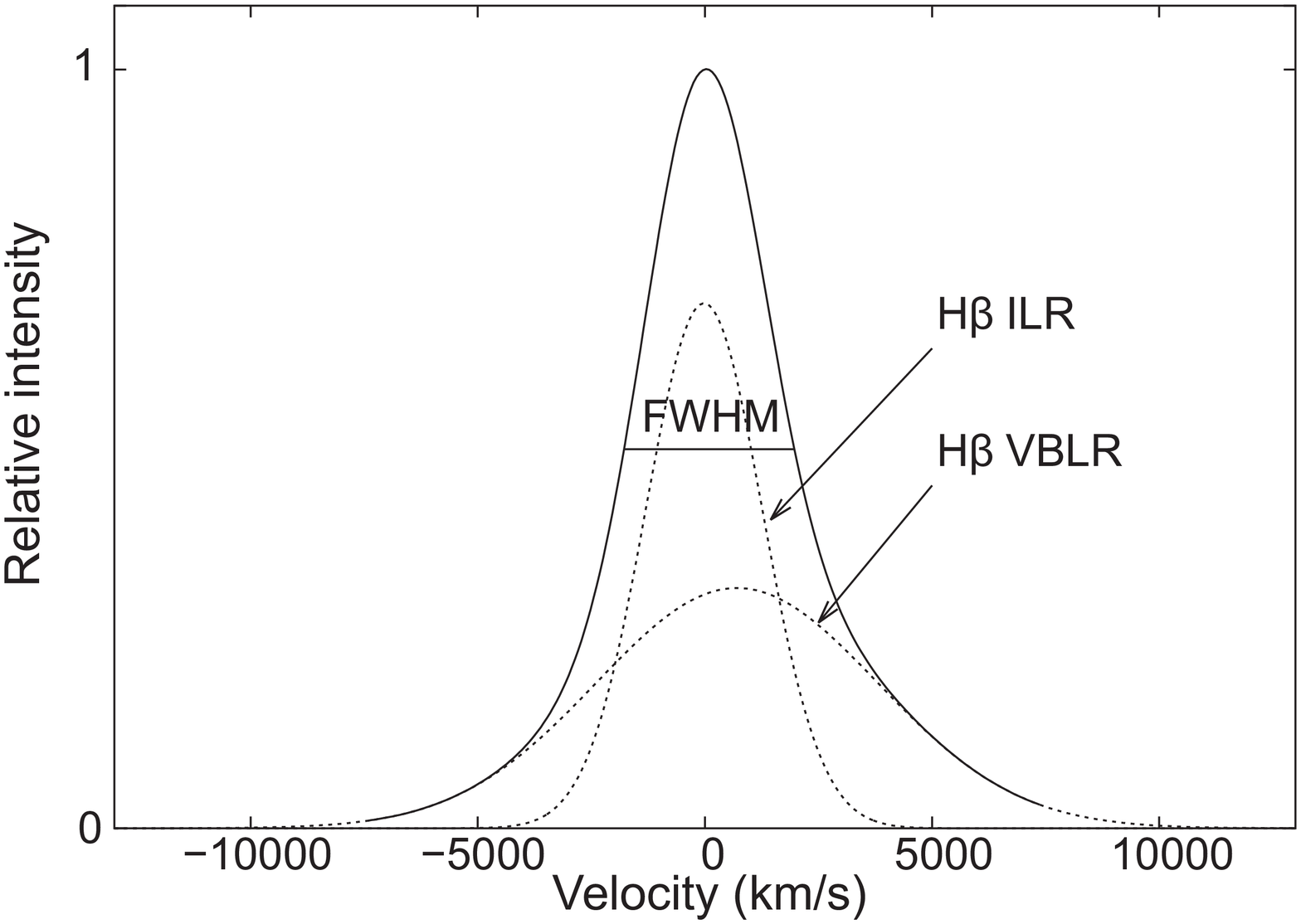}
\includegraphics[width=0.47\textwidth]{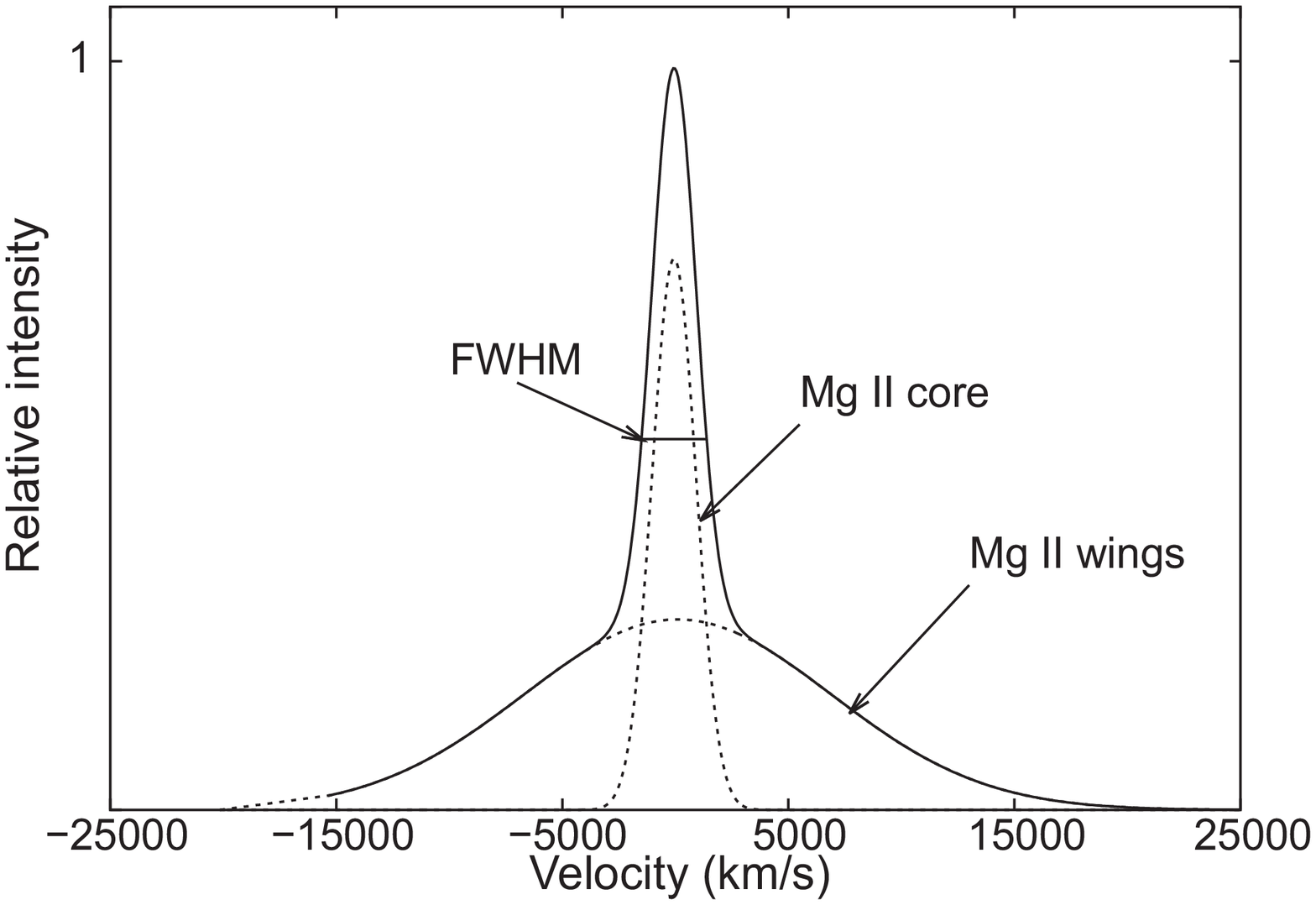}

\caption{Example of measuring FWHM of H$\beta$ (left) and \ion{Mg}{2} (right).
}
\label{4a}
\end{figure*}

We measure the Full Width at Half Maximum (FWHM) of broad lines, i.e. 
 Balmer lines (H$\beta$, H$\gamma$ and H$\delta$) and  the \ion{Mg}{2} line. For the Balmer lines, the FWHM is
measured for broad component (ILR + VBLR component), after subtraction of the narrow component.
Since \ion{Mg}{2} line has no a narrow component,
we measure the FWHM for the whole line (\ion{Mg}{2} core + \ion{Mg}{2} wings). 
The illustrations of determination of the  FWHM for H$\beta$ and \ion{Mg}{2} are 
shown in Fig \ref{4a}.

\section{Results}

\subsection{Kinematics of the \ion{Fe}{2} emission regions}

Kinematical properties of the lines (widths and shifts) reflect the motion of the emitting gas. The width of the 
line (or the line component) depends on random or gravitational bounded
motion of the emitting gas, while the shift is caused by the systemic motion of the
gas in an emission region. 
Therefore, similarities between kinematical properties of the different emission lines may indicate
a kinematical connection between their emission regions.

In order to investigate the kinematical connections between the UV and optical emission regions, we perform the correlations 
between the kinematical properties of the analyzed UV and optical emission lines. The widths and shifts of the 
lines and their components (represented by different Gaussians) are obtained from the best fit. 
As mentioned in Sec 2.1, in our fitting model we assume that all narrow lines (NLR component of Balmer lines and
[\ion{O}{3}] lines) are arising in the same emission region, and therefore have the same Doppler widths and shifts.

Similarly, it is assumed that the core components of all Balmer lines arise in ILR, and the wing components in VBLR,  and 
consequently they have the identical kinematical properties. 
Therefore, we now investigate possible correlations between kinematical parameters of the lines which are the free parameters, i.e. between Balmer 
components 
which arise in the NLR, ILR, VBLR, \ion{Mg}{2} core, \ion{Mg}{2} wings and the \ion{Fe}{2} lines in the UV and optical range.  The shifts are measured relative to the shift of the narrow lines ([\ion{O}{3}] 5007 \AA).

\begin{figure*}
\includegraphics[width=0.50\textwidth]{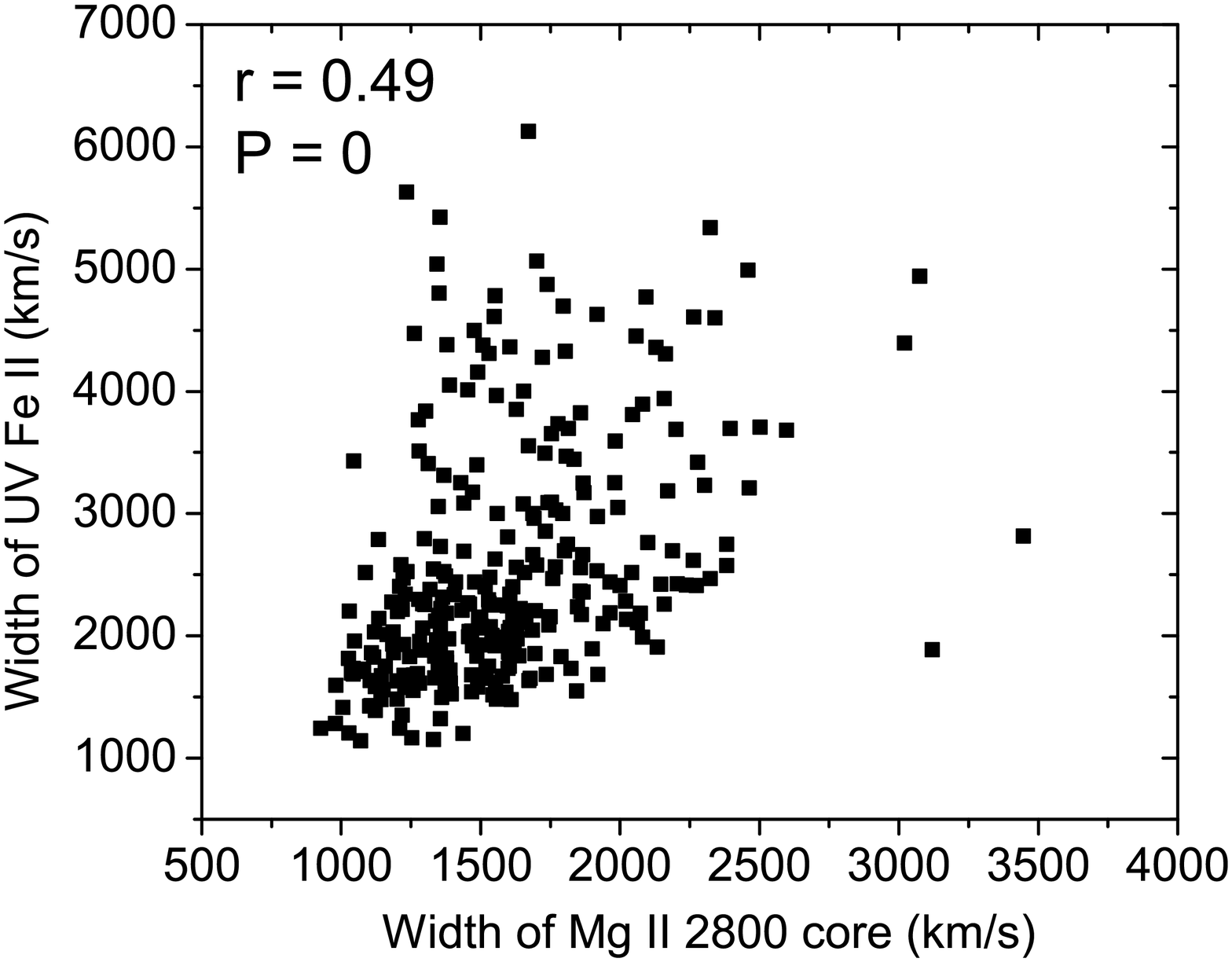}
\includegraphics[width=0.50\textwidth]{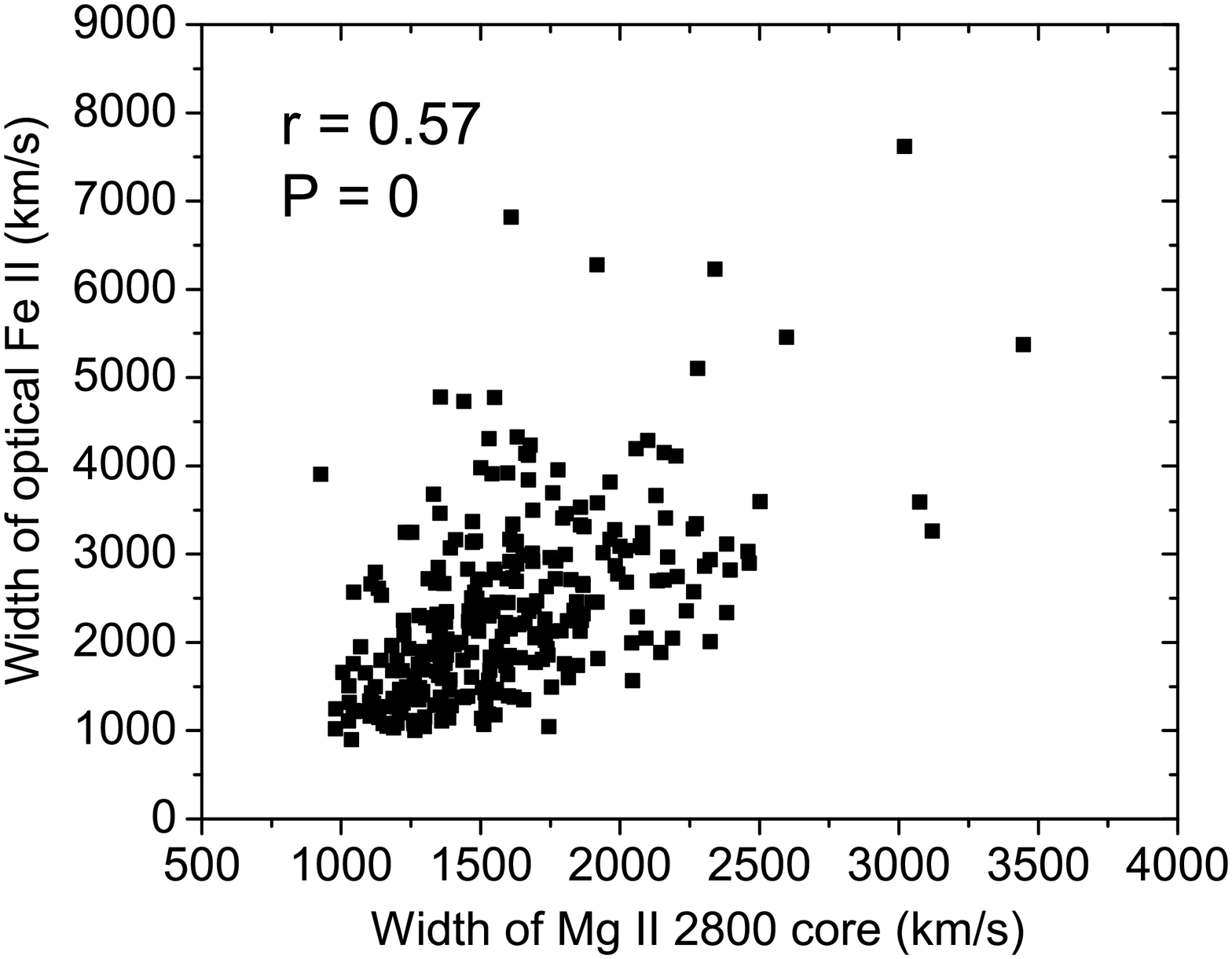}

\caption{Correlation between the Doppler widths of \ion{Mg}{2} 2800 core and UV \ion{Fe}{2} (left) and optical \ion{Fe}{2} (right).}
\label{5}
\end{figure*}

\begin{figure*}

\includegraphics[width=0.50\textwidth]{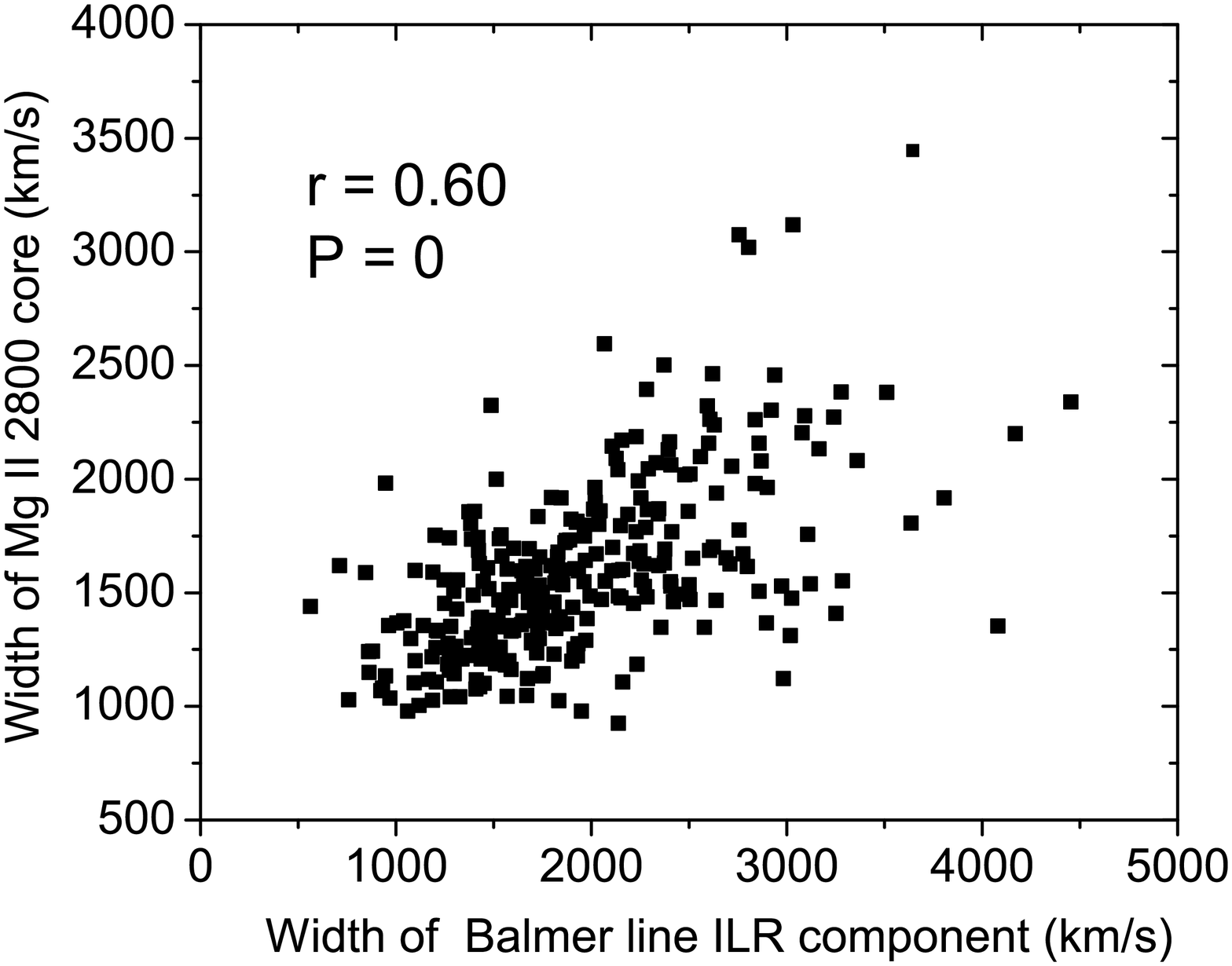}
\includegraphics[width=0.50\textwidth]{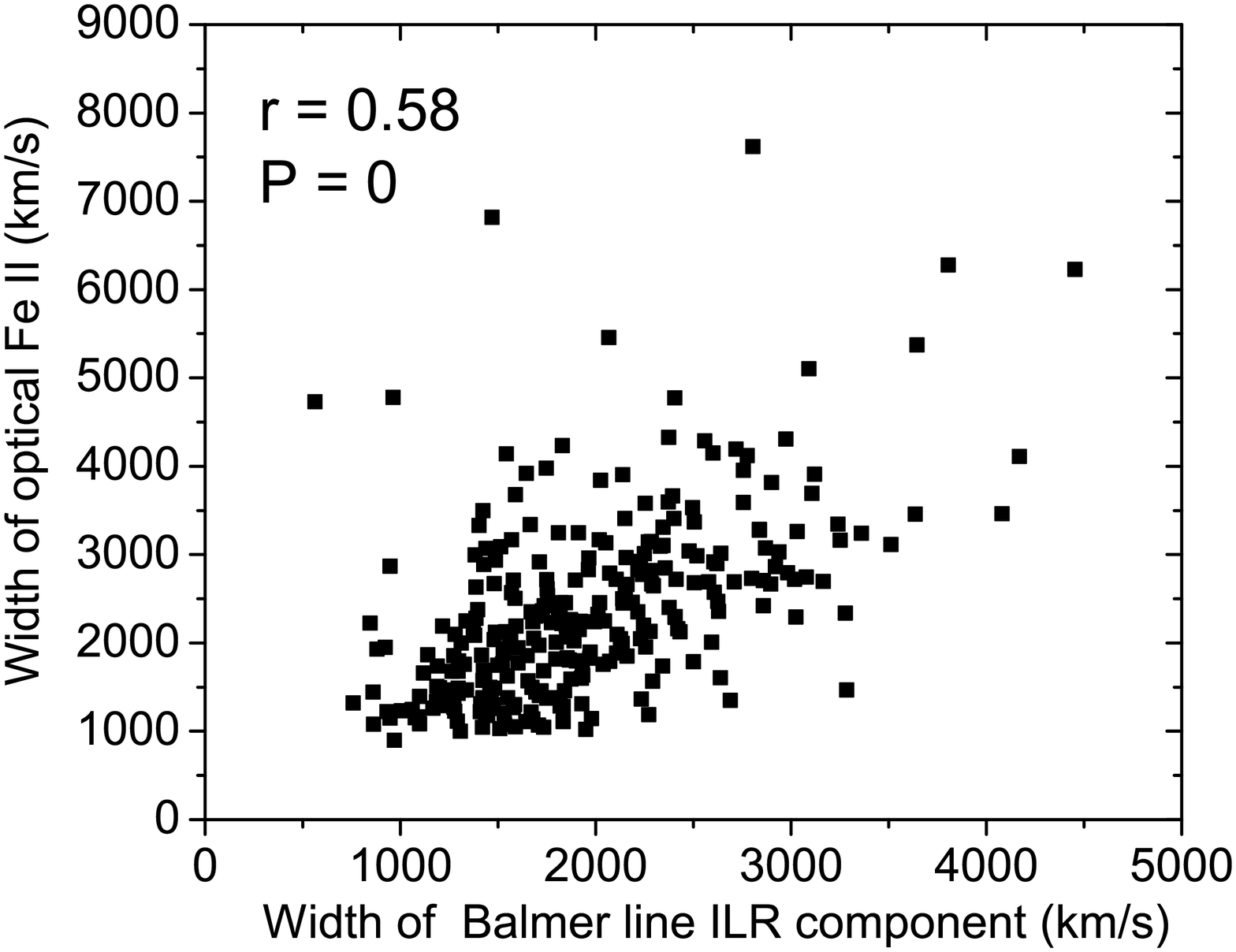}
\caption{Correlation between Doppler widths of the Balmer line ILR component and \ion{Mg}{2} 2800 core (left) and \ion{Fe}{2} optical (right).}
\label{6}
\end{figure*}

The correlations between the line widths and shifts are given in Table \ref{tbl-2} and Table \ref{tbl-3}, and the most significant are shown in 
Figs \ref{5} and \ref{6}. 
As it can be seen, the strongest kinematical connection is between the UV Fe II, optical Fe II, \ion{Mg}{2} core, and
Balmer lines core 
(ILR component). 
 
It is interesting that the  correlation between 
 the widths of the 
 UV \ion{Fe}{2} and optical \ion{Fe}{2} lines is weaker than correlation of their widths with some other lines from the 
 spectral range. 
 The strongest correlation of the UV \ion{Fe}{2} width is with the core of the \ion{Mg}{2} (r=0.49,
P=0; Fig \ref{5},
left), whereas its correlations with the widths of the optical \ion{Fe}{2} lines and Balmer line ILR component 
are slightly  less significant (r=0.39, P$\approx$2E-12).

The width of the optical \ion{Fe}{2}  lines has the most significant
correlation with the width of the \ion{Mg}{2} core as well
(r=0.57, P=0; Fig \ref{5}, right), and  with the  Balmer  ILR component (r=0.58,
P=0; Fig \ref{6}, right). 
The last correlation has was noted in  the previous work of \cite{ko2010}.

 We find that the correlation of the UV \ion{Fe}{2} and \ion{Fe}{2} optical widths are slightly higher with FWHMs of \ion{Mg}{2} and H$\beta$
(core + wings included), than with \ion{Mg}{2} core and H$\beta$ ILR components alone (see Table 2).  On the other hand, there is no a 
positive  correlation between the iron line widths and the wing component of \ion{Mg}{2} or H$\beta$:
there is only an anticorrelation with \ion{Mg}{2}  wings.

\begin{figure*}

\includegraphics[width=0.40\textwidth,angle=270]{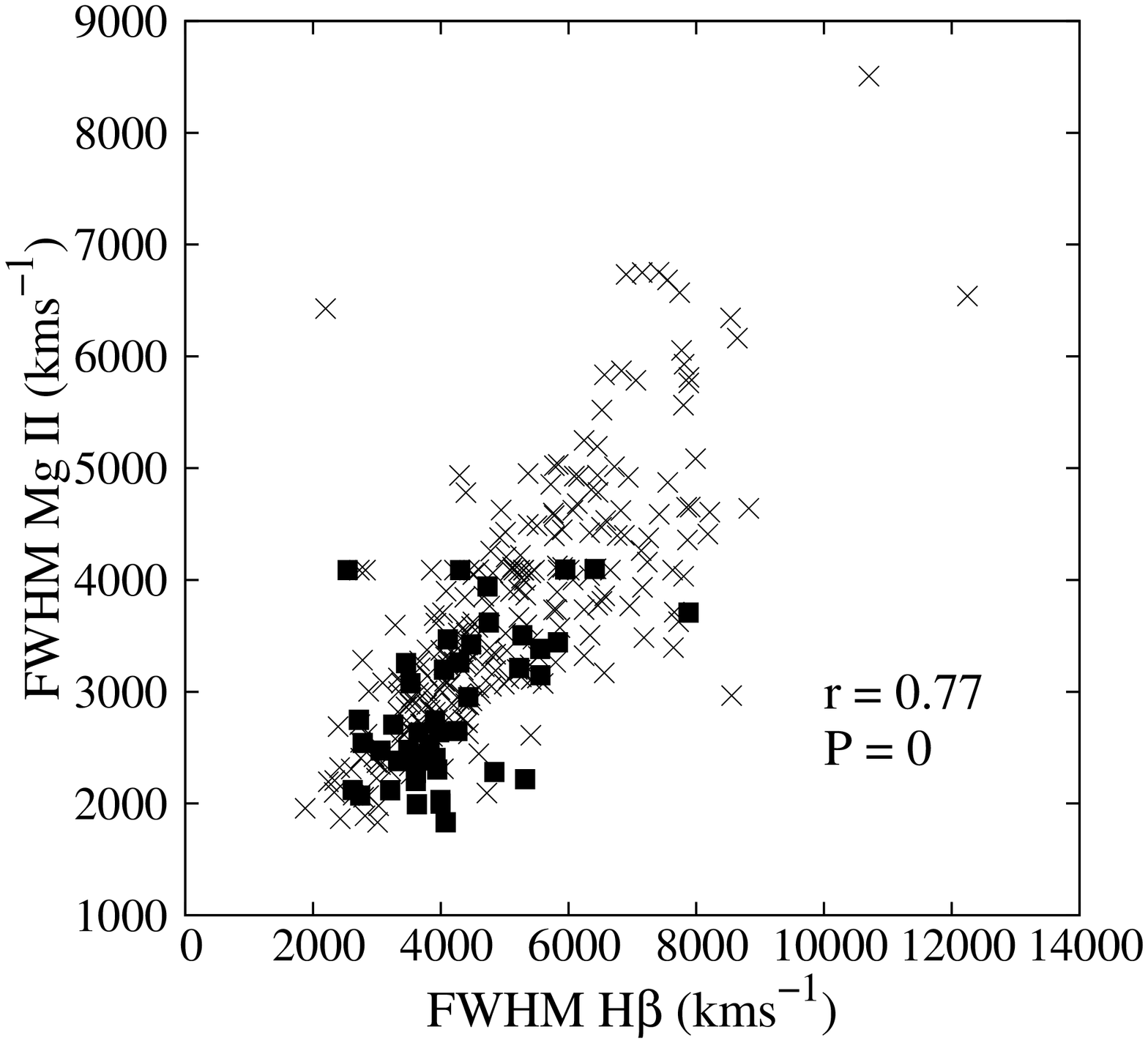}
\caption{Correlation between the FWHMs of \ion{Mg}{2} and H$\beta$. The fit with linear 
function f(x)=A*x+B, gives the values: A=0.504$\pm$0.026 and B=1086.09$\pm$135.2. Black squares: the objects with 
$\mathrm{log([O III]/H\beta}_{NLR}\mathrm{)<0.5}$ (with dominant starburst radiation), x-marks: the objects
with $\mathrm{log([O III]/H\beta}_{NLR}\mathrm{)>0.5}$ (pure AGNs).}
\label{6aaa}
\end{figure*}

The width of \ion{Mg}{2} line significantly correlates with the width of Balmer lines.
The correlation between FWHM H$\beta$ vs. FWHM \ion{Mg}{2} is r=0.77 and P=0 (see Fig. \ref{6aaa}).
As it can be seen, the objects
with $\mathrm{log([O III]/H\beta}_{NLR}\mathrm{)<0.5}$ (black squares  in Fig. \ref{6aaa}),
are located among the objects with smaller widths of H$\beta$ and \ion{Mg}{2}. 
The \ion{Mg}{2} core width correlates with the width of the Balmer line ILR components
(r=0.60, P=0; Fig \ref{6}, left),
while the width of the \ion{Mg}{2} wing component decreases as 
the widths of the Balmer line ILR, UV \ion{Fe}{2}, optical \ion{Fe}{2} and \ion{Mg}{2} core increase (see Table
\ref{tbl-2}). In addition, the wings of the \ion{Mg}{2} become narrower when this component is shifted to the red. The H$\beta$ ILR and \ion{Mg}{2} core are broader as the H$\beta$ VBLR component is shifted to the red.

Similarly as the widths, the shifts of the optical Fe II, UV Fe II, \ion{Mg}{2} core, and H$\beta$ ILR, are correlated (see Table
\ref{tbl-3}) and reflect a kinematical connection between their emission regions.
The shifts of the UV and optical \ion{Fe}{2} lines have the most significant correlation with the shift of the \ion{Mg}{2} core (r=0.48, 
P=0 for UV \ion{Fe}{2} and r=40, P=1.05E-12 for the optical Fe II), whereas the correlations with the H$\beta$ ILR are weaker. 
The strongest correlation is between the  shifts of the \ion{Mg}{2} core and Balmer line ILR (r=0.62, P=0).

\begin{figure*}

\includegraphics[width=0.50\textwidth]{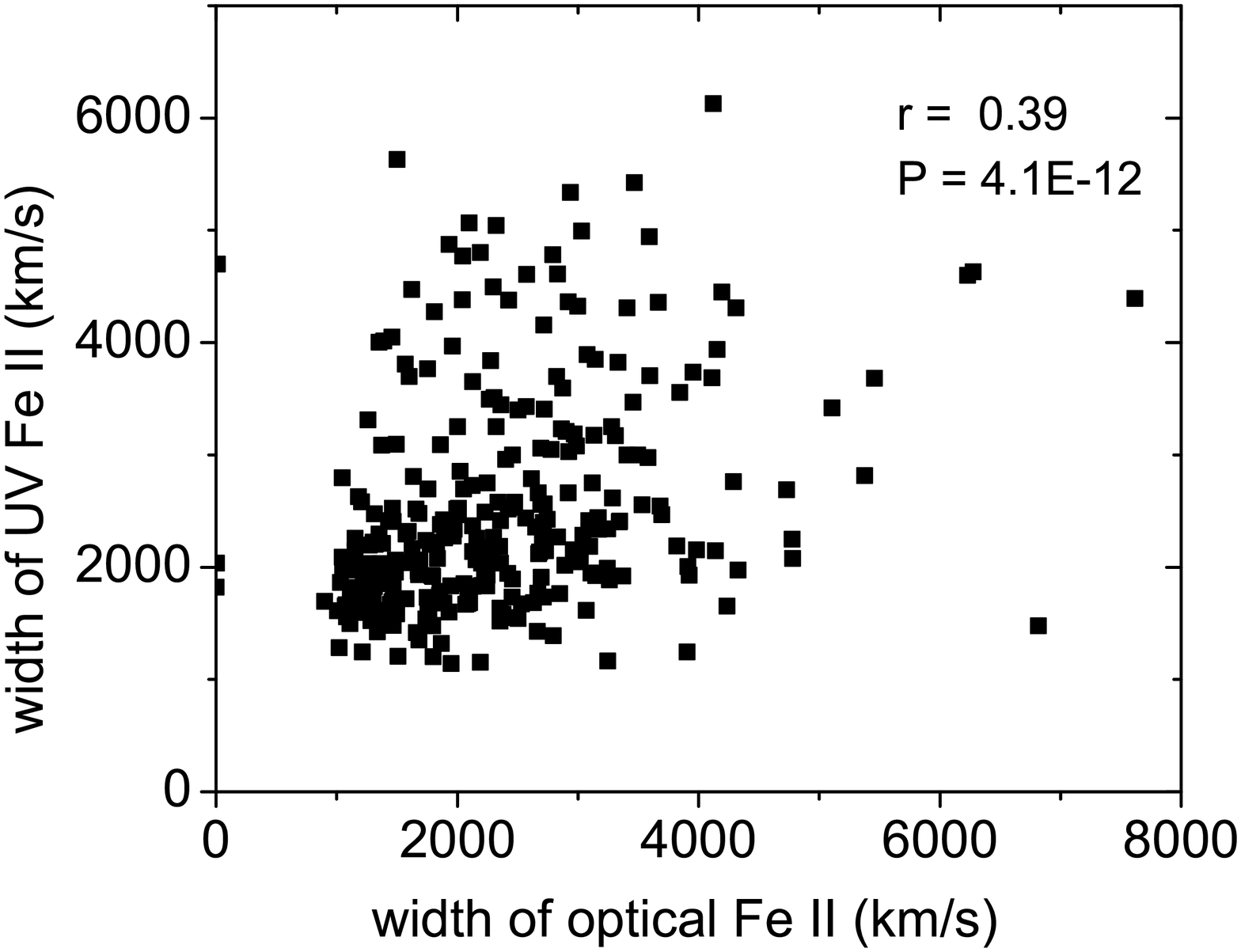}
\includegraphics[width=0.48\textwidth]{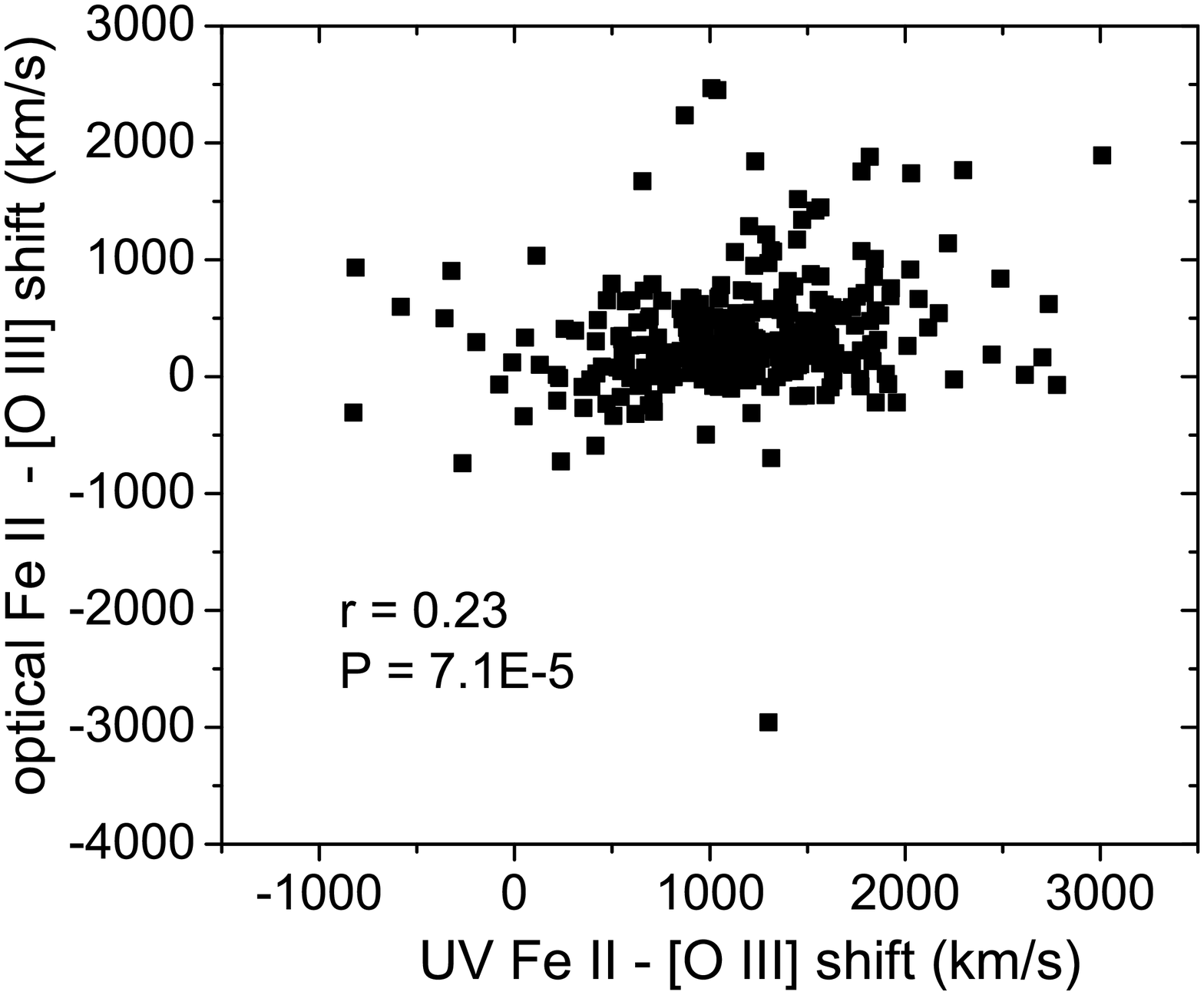}
\caption{Trend between the Doppler widths (left) and velocity shifts (right) of the optical and UV iron lines.}
\label{6a}
\end{figure*}

The relation between the  kinematical properties of the optical and UV iron lines is shown in
 Fig \ref{6a}. There are only weak trends between their widths, as well as between their shifts.

The average values for widths and shifts of analyzed lines and their components are given in Table \ref{tbl-4}. It can be
seen that the optical and UV \ion{Fe}{2}  lines have close average values for 
widths (optical Fe II: 2360 km s$^{-1}$, UV Fe II: 2530 km s$^{-1}$) with a very large dispersion.

\begin{figure*}

\includegraphics[width=0.35\textwidth,angle=270]{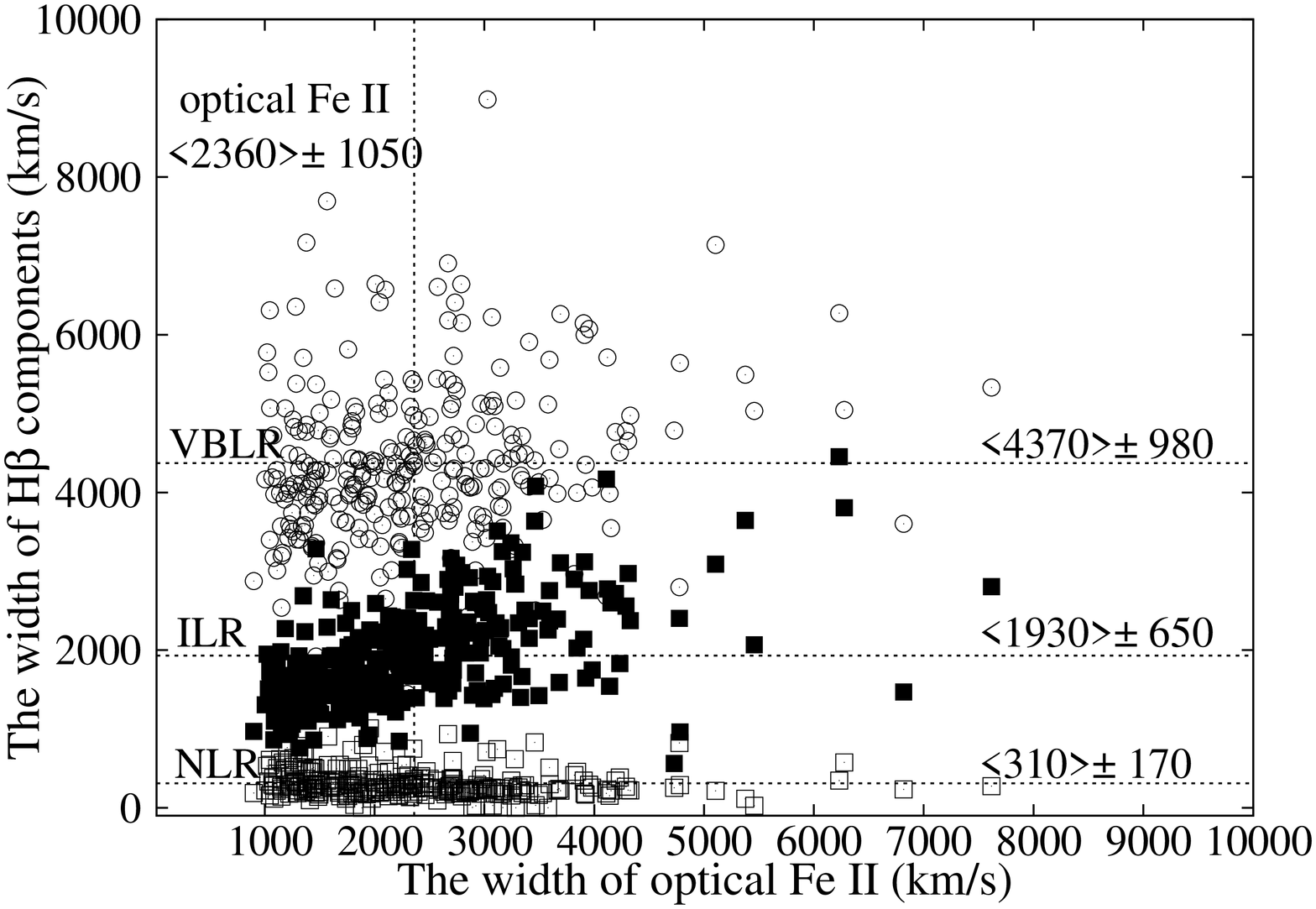}
\includegraphics[width=0.35\textwidth,angle=270]{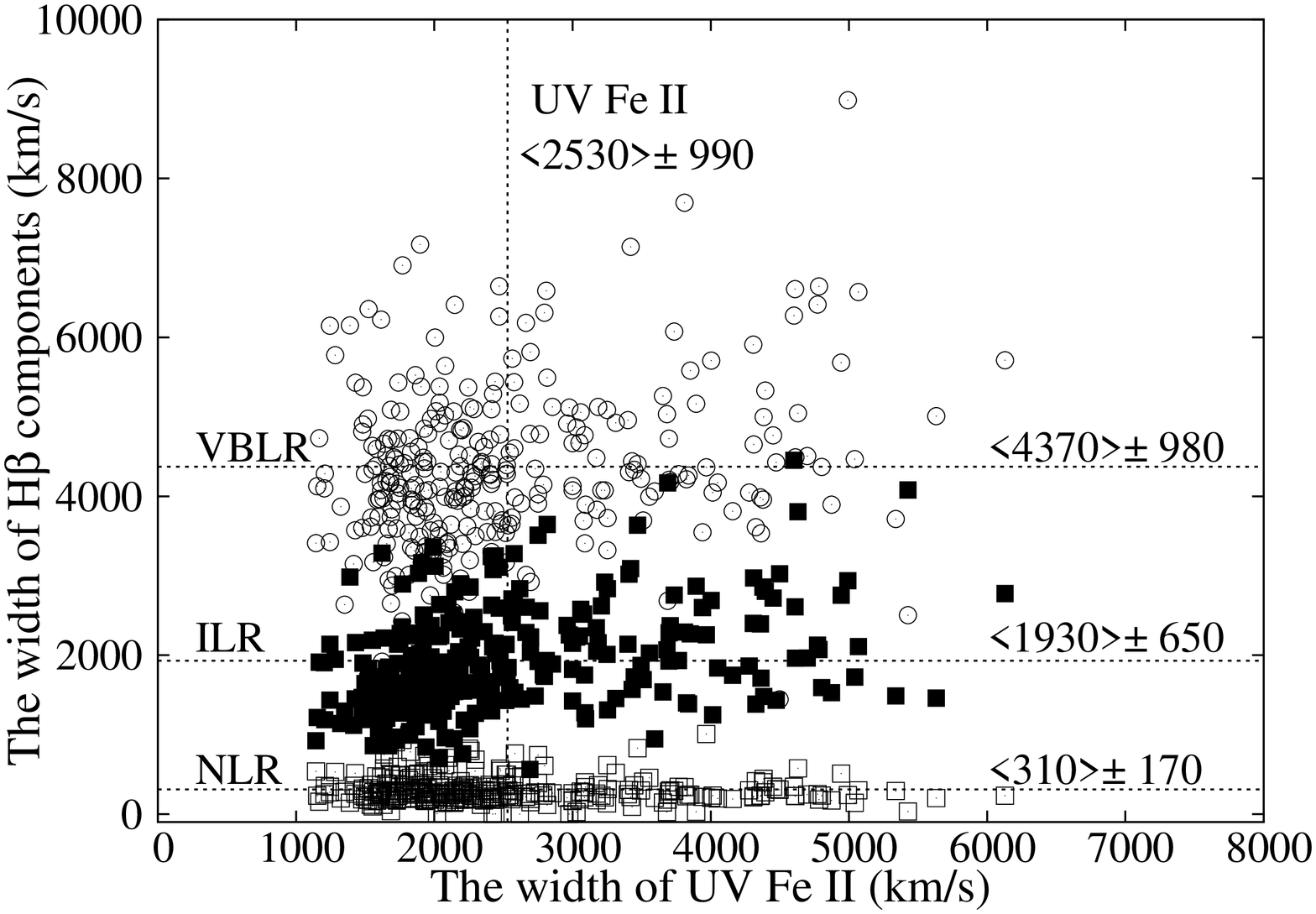}
\caption{ Comparison between the widths of the Balmer line components and the widths of optical (left) and UV (right) \ion{Fe}{2} lines.
The widths of the Balmer line components are presented on Y-axis as: NLR (white squares), ILR (black squares) and VBLR (circles).
The average values of widths are assigned with dotted line.}
\label{7}
\end{figure*}

\begin{figure*}

\includegraphics[width=0.35\textwidth,angle=270]{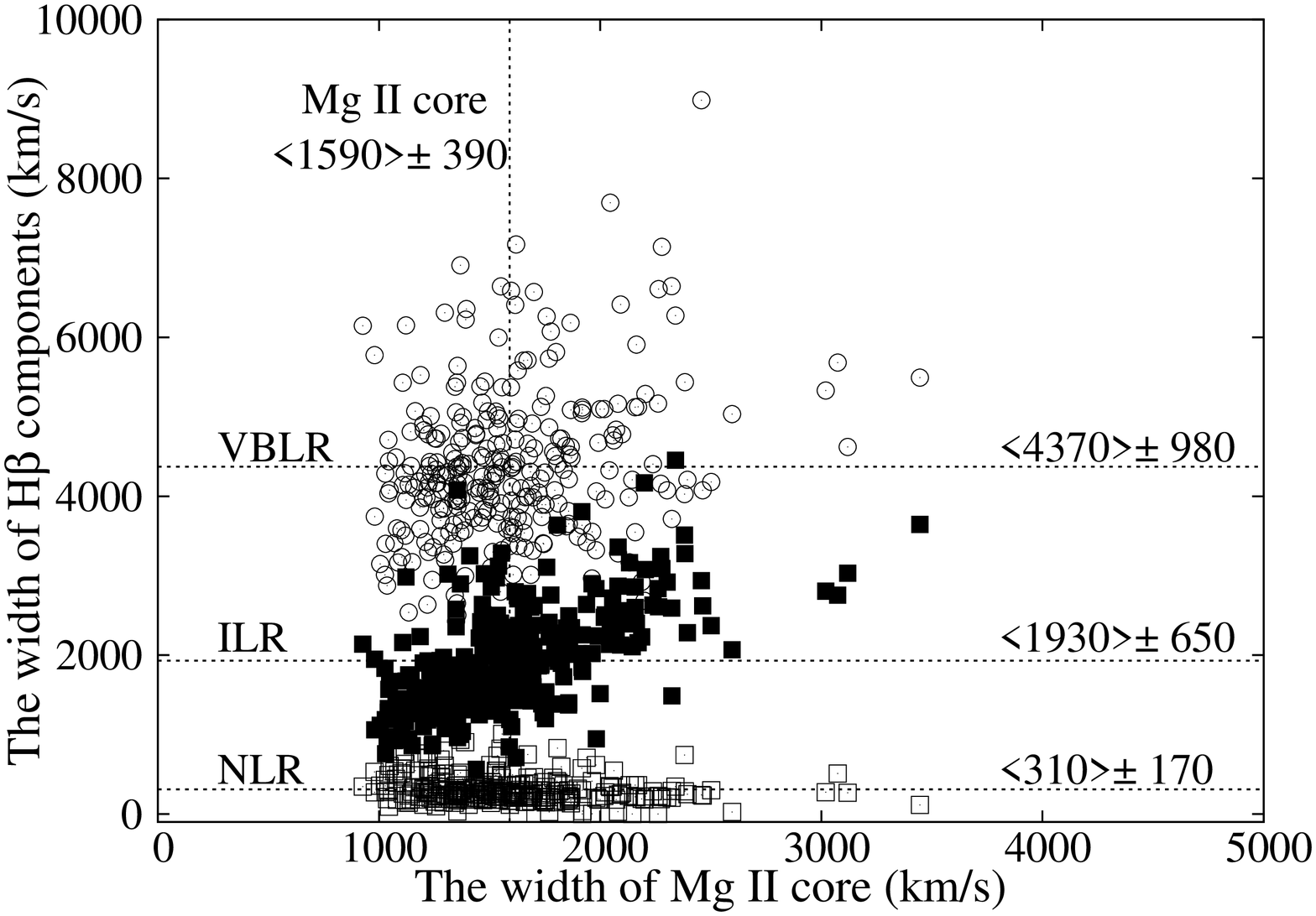}
\includegraphics[width=0.35\textwidth,angle=270]{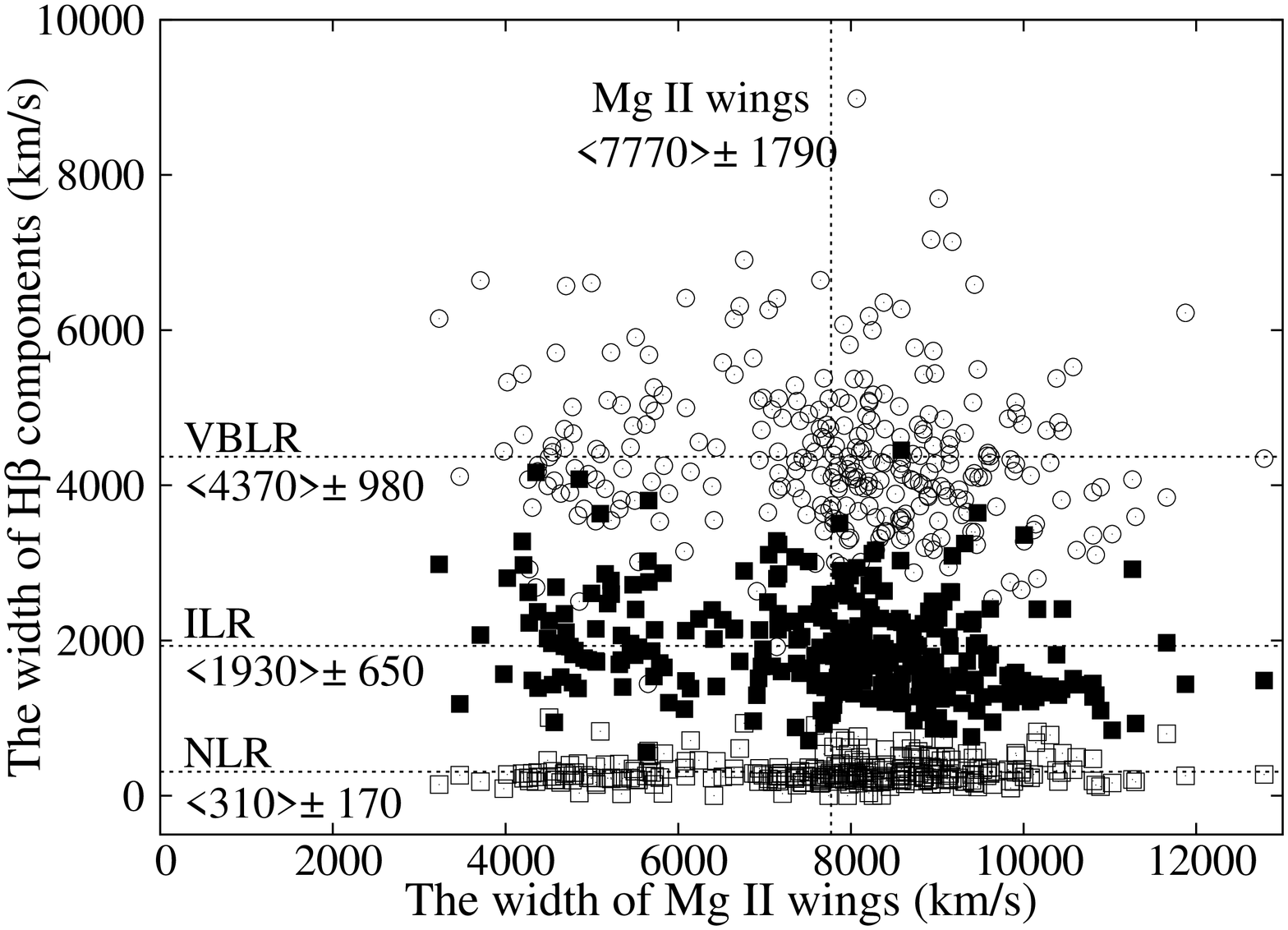}
\caption{Same as in previous Figure, just for the widths of \ion{Mg}{2} core (left) and \ion{Mg}{2} wings (right).}
\label{8}
\end{figure*}

 The majority of objects have the Doppler width of the optical 
\ion{Fe}{2} between 1000-3500 km s$^{-1}$ and of the UV \ion{Fe}{2} between 1500-3000 km s$^{-1}$. In Fig \ref{7}, 
the optical and UV \ion{Fe}{2} widths are
compared with the widths of Balmer line components, which originate from different line emission regions (NLR, ILR, VBLR), 
and the average values of the widths are assigned. The core of \ion{Mg}{2} lines and ILR component of Balmer lines have smaller average Doppler widths (\ion{Mg}{2} core: 
1590 km s$^{-1}$, ILR: 1930 km s$^{-1}$). The dispersion of the ILR widths is large as well, and most objects have an 
ILR width between 
1000-3000 km s$^{-1}$, whereas the dispersion of the \ion{Mg}{2} core is narrower and for 85\%
of objects the width is within range: 1000-2000 km s$^{-1}$. The comparison between the widths of the Balmer 
line components 
and \ion{Mg}{2} components are shown in Fig \ref{8}.

 The average  shift is significant only for the UV \ion{Fe}{2} lines (1150$\pm$580 km s$^{-1}$, see Table \ref{tbl-4}), which seems to be systematically redshifted relative to the narrow lines. All other analyzed lines (optical Fe II, \ion{Mg}{2} and H$\beta$ components)
have no significant average velocity shift.

\subsection{Correlations between the UV and optical emission line parameters}

A line EW reflects the emission line strength relative to the total continuum and it depends on a number of physical 
parameters of the emitting plasma, such as, e.g. electron density, temperature, the strength of the photoionizing 
flux, optical depth for the line, etc. We measure the EWs of broad and narrow lines from the spectral range and we find their averaged values for the sample, 
and for two subsamples with $\mathrm{log([O III]/H\beta}_{NLR}\mathrm{)<0.5}$ (SB dominant) and $\mathrm{log([O III]/H\beta}_{NLR}\mathrm{)>0.5}$ (AGN dominant). The results are shown in 
Table \ref{tbl-nova1}.
 As it can be seen, there are differences
between AGN and SB dominant objects, and the biggest one is for the average EW of the optical \ion{Fe}{2} and [\ion{O}{3}] lines. The
SB dominate subsample has significantly higher EW of optical \ion{Fe}{2} and weaker EW [\ion{O}{3}] than the AGN dominant subsample.

We explore the correlations between the EWs of the optical and UV lines and their components (see Table  \ref{tbl-7}).  We can 	
summarize the correlations given in Table \ref{tbl-7} as:
(i) there is no any correlation between the EWs of the optical and UV \ion{Fe}{2} lines; (ii) the EW \ion{Fe}{2} UV correlates only with the EW Mg II$_{total}$; (iii) the EW \ion{Fe}{2} optical shows only anticorrelation with EW [\ion{O}{3}] \citep{bg92, ko2010}; 
(iv) the EW [\ion{O}{3}] correlates with EWs of all analyzed lines and line components (broad and narrow), except with the EW \ion{Fe}{2} UV and anticorrelates only with
the EW \ion{Fe}{2} optical; (v) there is a correlation among EWs of all narrow lines; (vi) the EW \ion{Mg}{2} correlates with EWs of all analyzed broad lines except with EW \ion{Fe}{2} optical.

\begin{figure*}

\includegraphics[width=0.50\textwidth,angle=270]{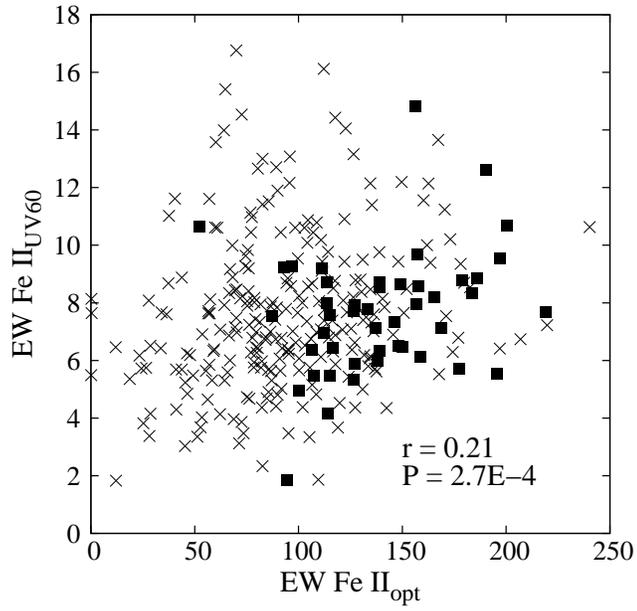}
\caption{Correlation between the EWs of optical and UV Fe II.  Black squares: the objects with 
$\mathrm{log([O III]/H\beta}_{NLR}\mathrm{)<0.5}$, x-marks: the objects with $\mathrm{log([O III]/H\beta}_{NLR}\mathrm{)>0.5}$.   }
\label{13aa}
\end{figure*}

There is no correlation between the EWs of the optical and UV \ion{Fe}{2} lines. The plot between the optical \ion{Fe}{2} and multiplet 60 of the UV \ion{Fe}{2} is shown in Fig \ref{13aa}.
 The multiplet 60 of the UV \ion{Fe}{2} was chosen because it is well defined feature at $\sim$2950 \AA \ and it does not overlap with extended \ion{Mg}{2} wings.
It can be seen that objects with a dominant starburst radiation (black squares  in Fig \ref{13aa}) generally have 
a strong optical \ion{Fe}{2} emission, while such trend cannot be seen for the UV \ion{Fe}{2} lines.

\begin{figure*}

\includegraphics[width=0.37\textwidth,angle=270]{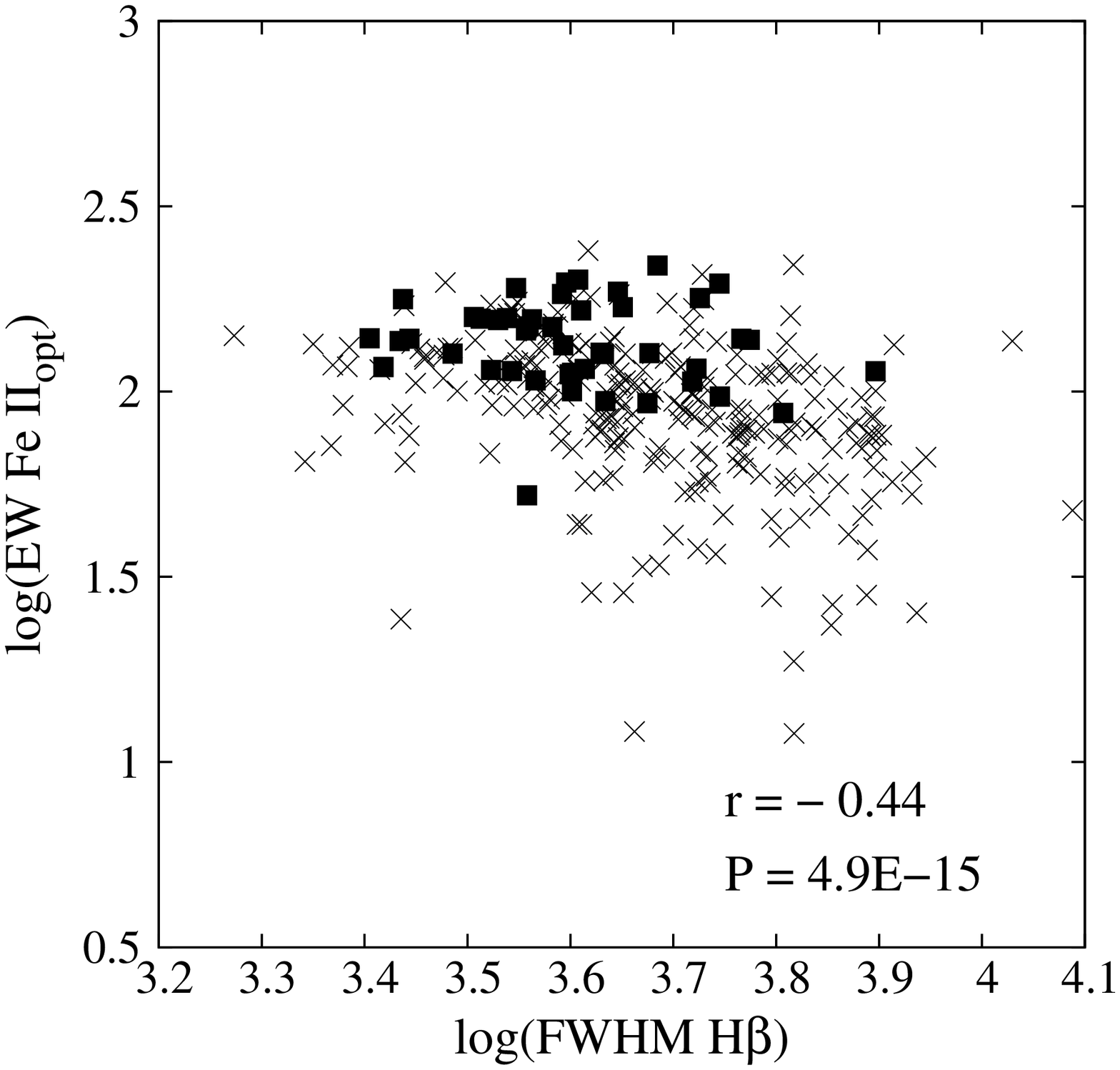}
\includegraphics[width=0.37\textwidth,angle=270]{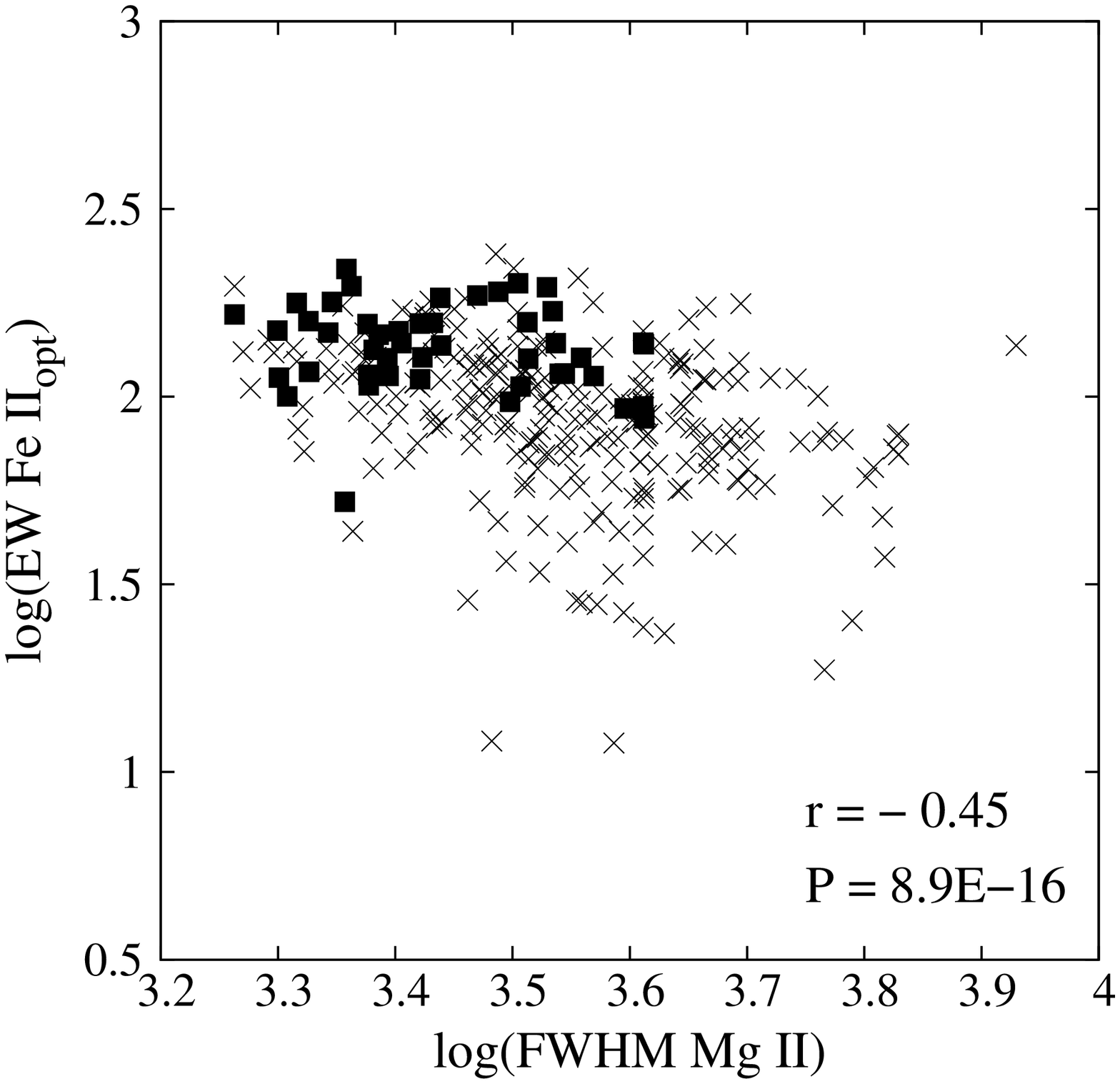}

\caption{Correlations between EW of optical \ion{Fe}{2} and FWHM of the broad H$\beta$  (left) and FWHM of \ion{Mg}{2} (right). 
Black squares: the objects with 
$\mathrm{log([O III]/H\beta}_{NLR}\mathrm{)<0.5}$, x-marks: the objects with $\mathrm{log([O III]/H\beta}_{NLR}\mathrm{)>0.5}$. }
\label{14}
\end{figure*}

 The correlations between the widths and EWs of the considered emission lines  are shown in Table \ref{tbl-9}. The most 
 interesting 
correlations in this table are those with the EW of the optical \ion{Fe}{2} and the line widths. While EWs of the Fe II$_{UV 60}$, H$\beta$ 
NLR and \ion{Mg}{2} line increase, their line widths increase as well; however, the opposite happens for the EW of the optical \ion{Fe}{2} lines: the EW Fe
II optical increases as its width decreases. Moreover, EW Fe II$_{optical}$ anticorrelates with the widths of all analyzed broad lines, 
especially with FWHM H$\beta$ and FWHM \ion{Mg}{2} (see Fig \ref{14}). 
 The objects with dominant SB emission (black squares in Fig. \ref{14}), as expected, have a strong optical \ion{Fe}{2} emission
and narrower H$\beta_{broad}$ and \ion{Mg}{2} lines.

\subsubsection{The flux ratios of the emission lines}

We explore the correlations between the line properties and flux ratios
$\mathrm{Fe II}_{opt}/\mathrm{Fe II}_{UV 60}$,
$\mathrm{Fe II}_{opt}/\mathrm{Mg II}$, $\mathrm{Fe II}_{UV}/\mathrm{Mg II}$, $\mathrm{Mg II}/\mathrm{H\beta_{broad}}$, 
$\mathrm{[O III]_{5007}}/\mathrm{H\beta_{NLR}}$, $\mathrm{H\beta}_{broad}/\mathrm{H\gamma}_{broad}$
(see Table \ref{tbl-8}). These flux ratios are chosen because they may be indicators of some physical conditions  or the abundance in the emission line regions. 

Different models of the iron emission predict that the ratio of $\mathrm{Fe II}_{opt}/\mathrm{Fe II}_{UV}$ could be an indicator of the
 column density, due the atomic properties of the iron ion \citep[see][]{jo1987, sa2011}.
 \citet{ve2003} and  \citet{sa2011}
 found that this ratio depends on microturbulence, i.e.
the increase of the $\mathrm{Fe II}_{opt}/\mathrm{Fe II}_{UV}$ ratio reflects the increase of the column density and the 
decrease of the microturbulence in the emitting gas.
It has been found that this ratio anticorrelates with FWHM \ion{Mg}{2} \citep{ts2006} and FWHM H$\beta$ \citep{do2011},
but correlates with the 
Eddington ratio \citep{sa2011, do2011}.

 Table \ref{tbl-8} shows several correlations between this ratio and the different line properties.
We found that the ratio  $\mathrm{Fe II}_{opt}/\mathrm{Fe II}_{UV 60}$ increases as: 
(i) the widths of all analyzed broad lines decrease (H$\beta$, Mg II, optical and UV Fe II);
(ii) the EW \ion{Mg}{2} decreases.

\begin{figure*}

\includegraphics[width=0.37\textwidth,angle=270]{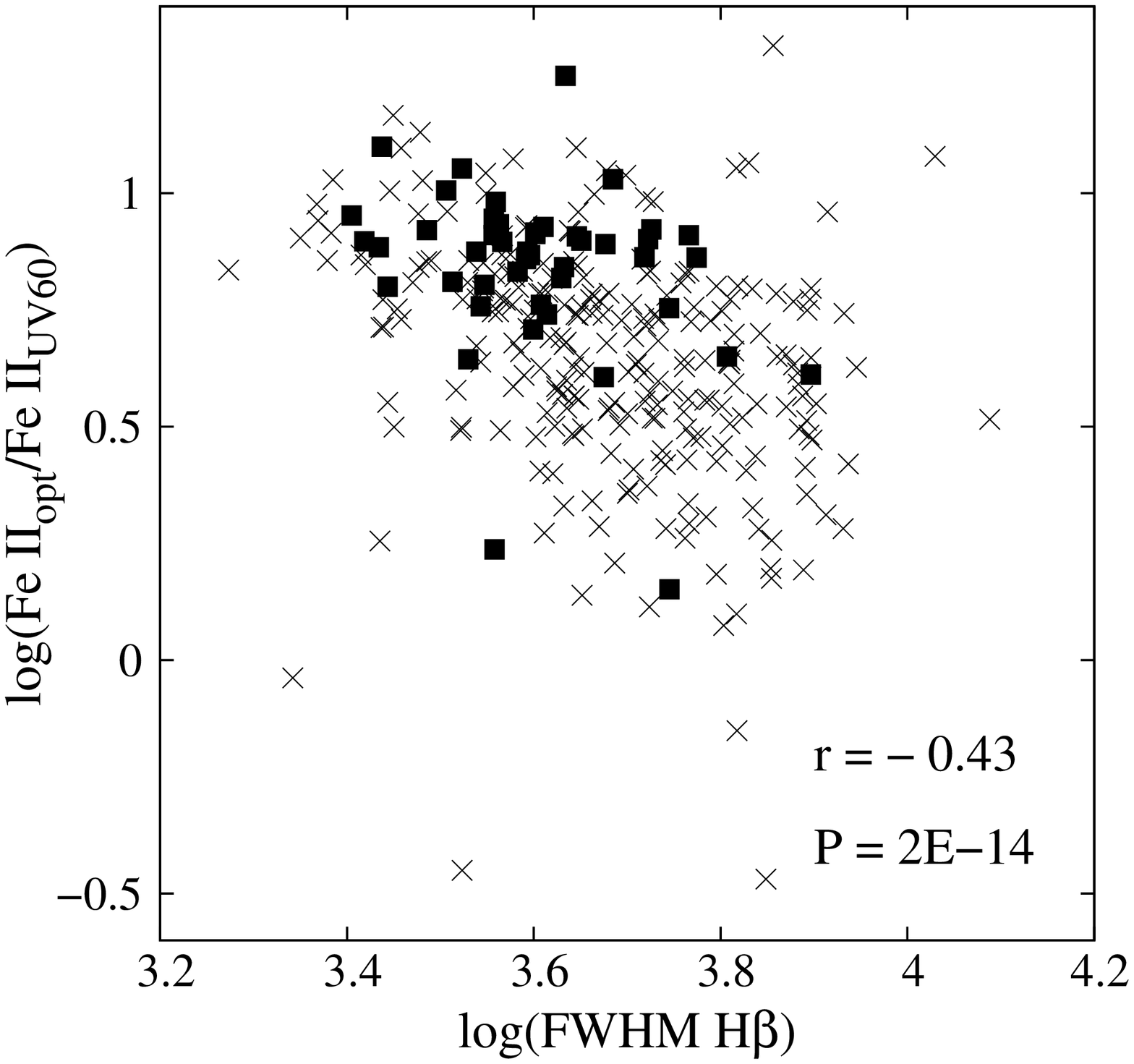}
\includegraphics[width=0.37\textwidth,angle=270]{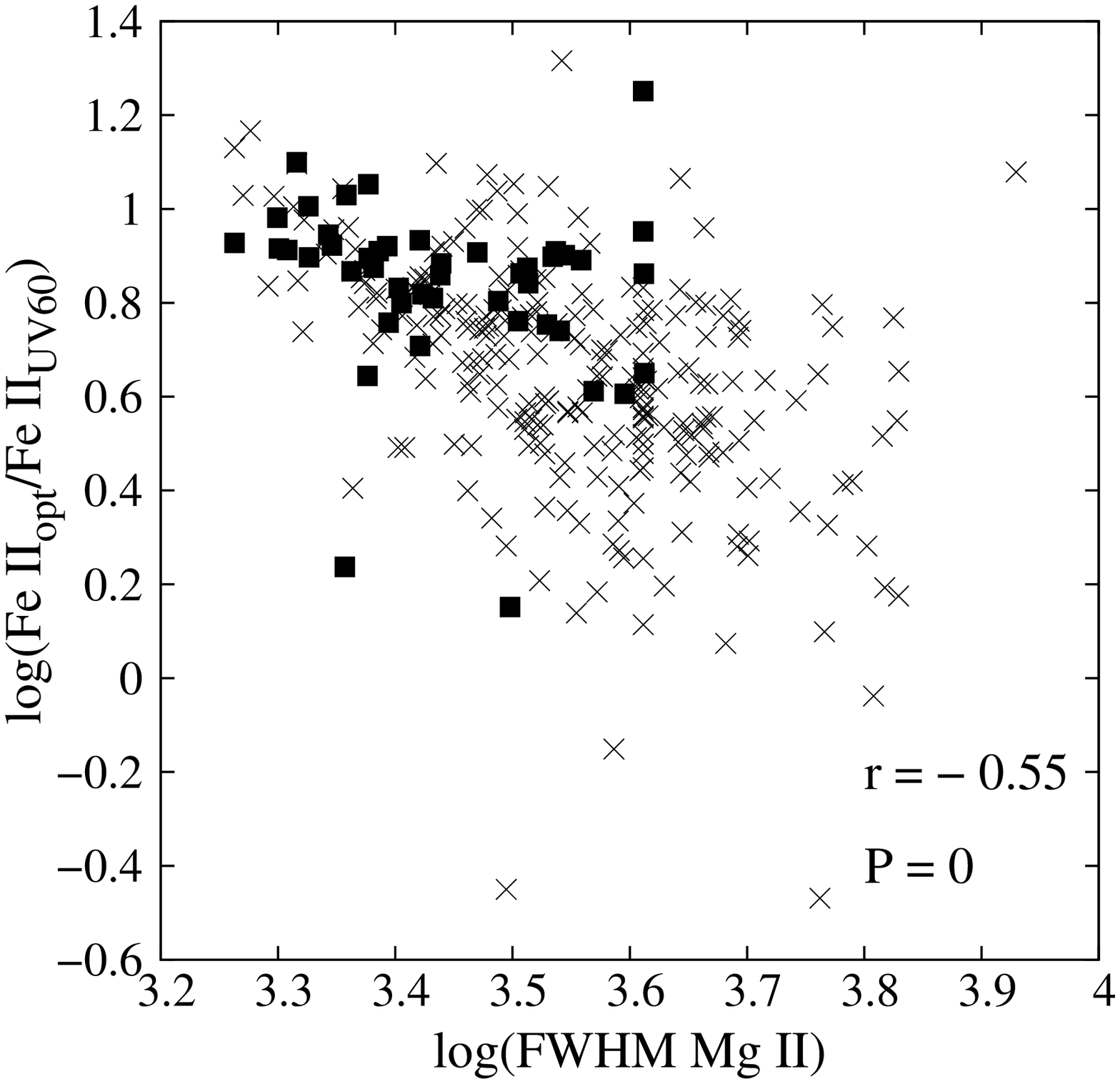}

\caption{Correlations between the $\mathrm{Fe II}_{opt}/\mathrm{Fe II}_{UV 60}$ and FWHM H$\beta$ (left) and  \ion{Mg}{2} width (right).  
Black squares: the objects with 
$\mathrm{log([O III]/H\beta}_{NLR}\mathrm{)<0.5}$, x-marks: the objects with $\mathrm{log([O III]/H\beta}_{NLR}\mathrm{)>0.5}$.}
\label{13}
\end{figure*}

 The correlations between the $\mathrm{Fe II}_{opt}/\mathrm{Fe II}_{UV 60}$ ratio and FWHMs of H$\beta$ and \ion{Mg}{2} are presented in Fig \ref{13}.
Note that SB dominant objects (assigned with black squares) are in the upper part of graphs among the objects with a high
$\mathrm{Fe II}_{opt}/\mathrm{Fe II}_{UV 60}$ ratio and small FWHM H$\beta$ and FWHM \ion{Mg}{2} widths.

 The ratio $\mathrm{Fe II}/\mathrm{Mg II}$ is considered by several authors to be a cosmological abundance indicator 
 \citep{hf1993, mr2001, ve2003, do2011, dr2011}. Until now, no relation has been found between 
this ratio and cosmological redshift, which may be a consequence of the additional influence of the physics of the emission region to this 
ratio. The model of \citet{ve2003} predicts that the
$\mathrm{Fe II}_{UV}/\mathrm{Mg II}$ ratio is sensitive on microturbulence. Similar to the $\mathrm{Fe II}_{opt}/\mathrm{Fe II}_{UV 60}$ ratio, the
$\mathrm{Fe II}_{opt}/\mathrm{Mg II}$ increases as the
FWHM \ion{Mg}{2} and FWHM H$\beta$ decrease \citep{ts2006, do2011}, and Eddington ratio increases  \citep{do2011}. 

\begin{figure*}

\includegraphics[width=0.37\textwidth,angle=270]{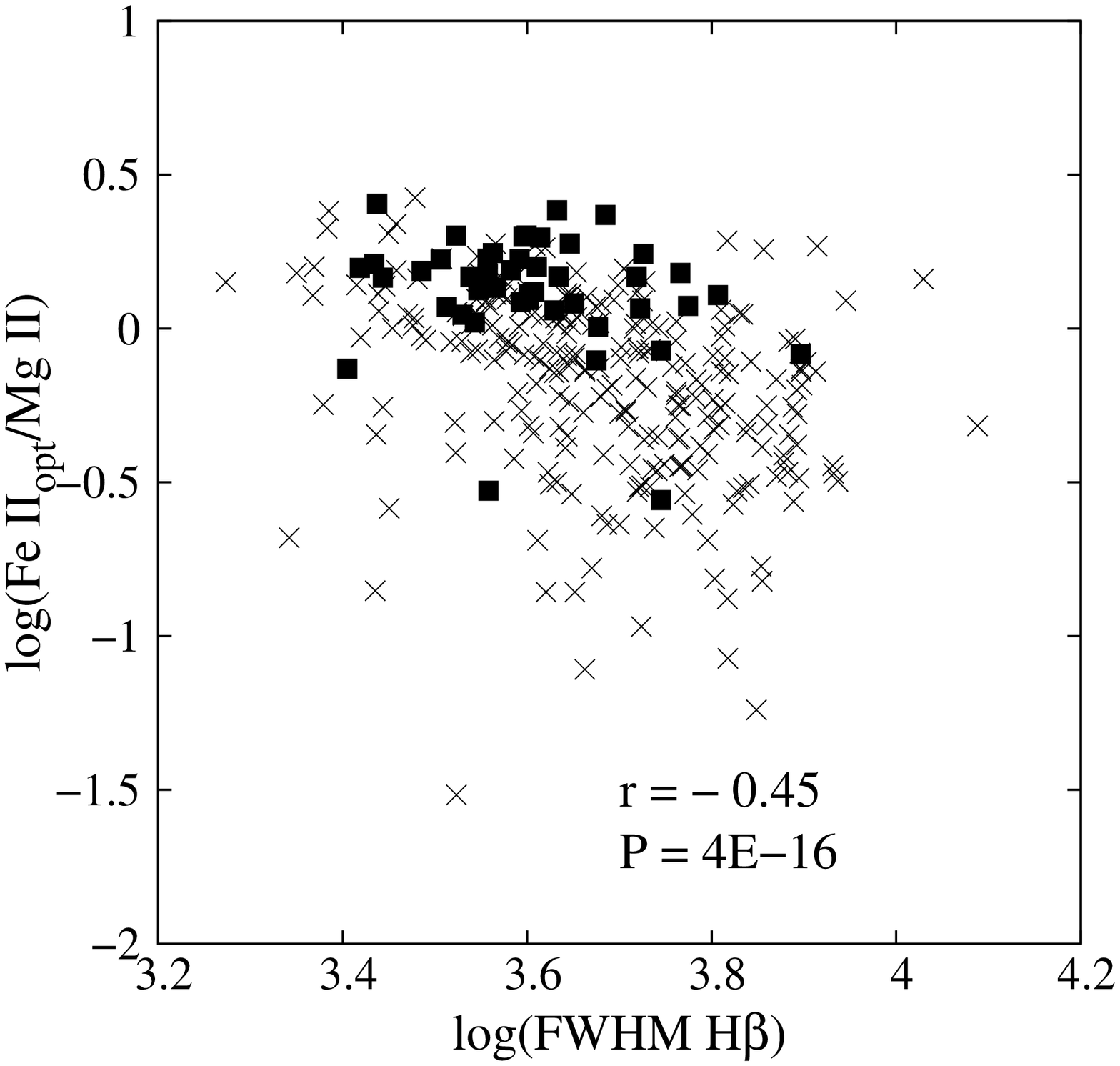}
\includegraphics[width=0.37\textwidth,angle=270]{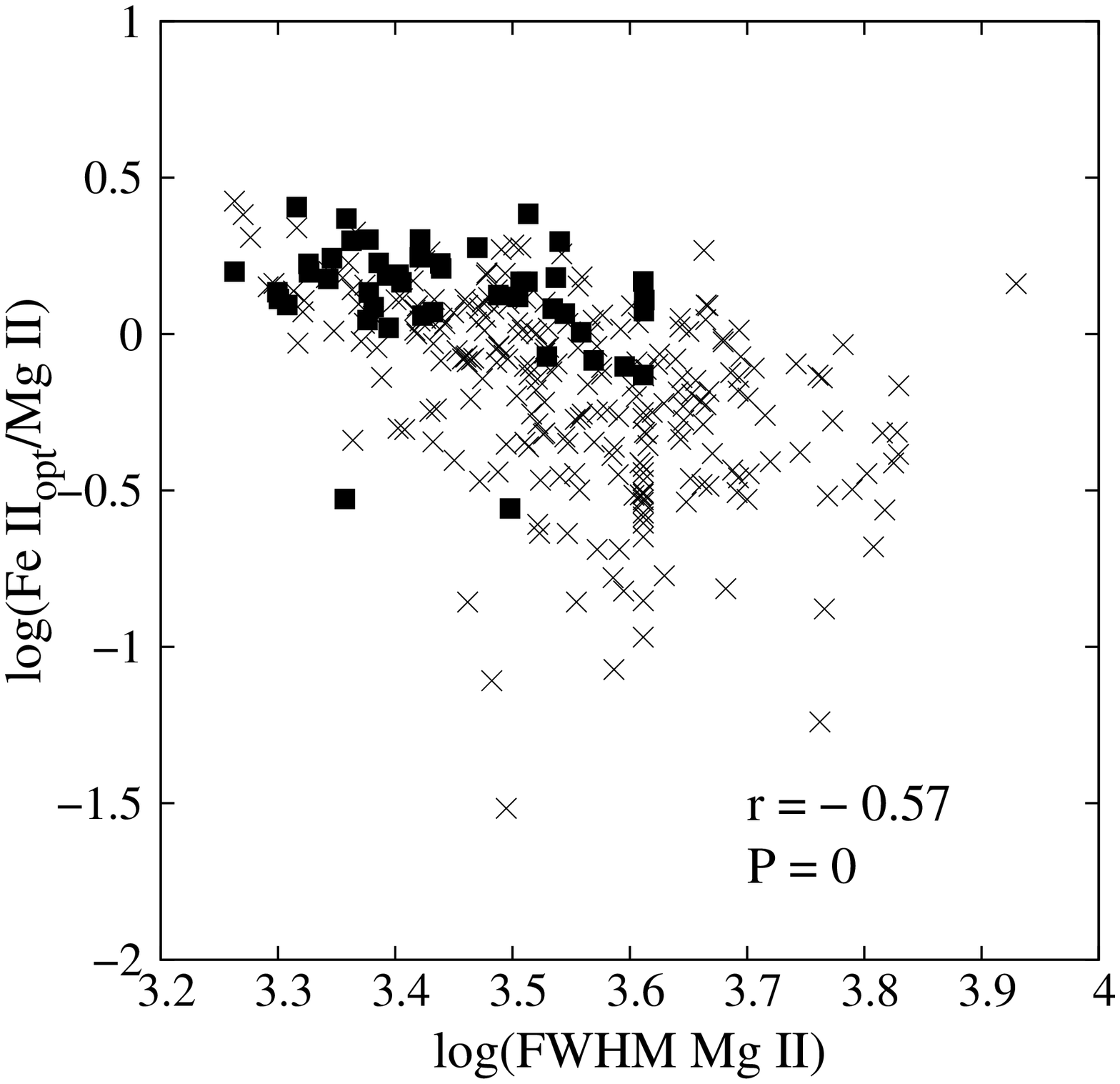}
\caption{Same but for the $\mathrm{Fe II}_{opt}/\mathrm{Mg II}$ ratio.  Black squares: the objects with 
$\mathrm{log([O III]/H\beta}_{NLR}\mathrm{)<0.5}$, x-marks: the objects with $\mathrm{log([O III]/H\beta}_{NLR}\mathrm{)>0.5}$.}
\label{14a}
\end{figure*}

 In our sample we analyze the both ratios, $\mathrm{Fe II}_{opt}/\mathrm{Mg II}$ and $\mathrm{Fe II}_{UV}/\mathrm{Mg II}$, and we find that correlations with some spectral properties are different for these two ratios.
 The $\mathrm{Fe II}_{opt}/\mathrm{Mg II}$ ratio increases as the widths of the broad lines (except \ion{Fe}{2} UV) decrease, while the ratio $\mathrm{Fe II}_{UV}/\mathrm{Mg II}$ does
not correlate with these 
properties. Both ratios ($\mathrm{Fe II}_{opt}/\mathrm{Mg II}$ and $\mathrm{Fe II}_{UV}/\mathrm{Mg II}$) anticorrelate with EW [\ion{O}{3}] 
and EW H$\beta_{broad}$. The correlation of the $\mathrm{Fe II}_{opt}/\mathrm{Mg II}$
and the widths of  H$\beta_{broad}$ and \ion{Mg}{2} are presented in Fig \ref{14a}. The starburst dominant objects (black 
squares in Fig \ref{14a}) have a high ratio of $\mathrm{Fe II}_{opt}/\mathrm{Mg II}$.

The ratio $\mathrm{Mg II}/\mathrm{H\beta_{broad}}$ may be taken as an  indicator of the  element abundance in AGNs.
This ratio correlates with the width of the \ion{Mg}{2} and with EW $\mathrm{Fe II}_{UV}$, that shows how beside the abundance,
other effects can affect on the $\mathrm{Mg II}/\mathrm{H\beta_{broad}}$ ratio.

Finally, the ratio $\mathrm{[O III]_{5007}}/\mathrm{H\beta_{NLR}}$ is assumed to be an indicator of the starburst activity. 
Namely, the ratios of some narrow lines reflect the shape of ionizing 
continuum, i.e. whether the ionization source is the accretion disc around black hole or hot, young stars. 
This fact is used in the construction of the diagnostic diagrams based on the narrow line ratios
\citep[see][]{bpt1981,vo1987}. \citet{pk11} found that the $\mathrm{[O III]}/\mathrm{H\beta_{NLR}}$ ratio 
(which is usually used as one axis in diagnostic diagrams)
could be used as an approximate indicator of the presence or absence of the starburst activity in AGN spectra. 

We analyze the correlations between this ratio and other spectral properties in our sample (see Table \ref{tbl-8}), and find
that as $\mathrm{[O III]}/\mathrm{H\beta_{NLR}}$ ratio increases (which indicate smaller contribution of SB fraction):
(i) the narrow lines become narrower;
(ii) the broad lines become broader;
(iii) the EWs of [O III], H$\beta_{broad}$ and \ion{Mg}{2} increase;
(iv) the EW of Fe II$_{opt}$ decreases.

On the other hand, the objects that may have significant SB activity near  an AGN
(which is reflected as decrease of this ratio), have a smaller difference in the width of the narrow and 
broad lines, stronger Fe II$_{opt}$ lines, and weaker [O III], H$\beta_{broad}$ and Mg II.
No correlation is seen for the EW of Fe II$_{uv}$.

\begin{figure*}

\includegraphics[width=0.40\textwidth,angle=270]{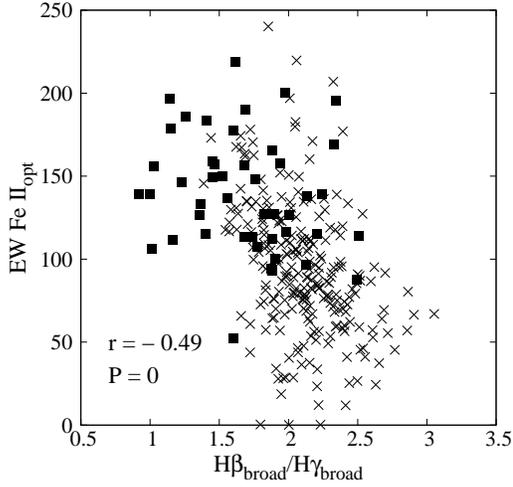}
\caption{Correlation between $\frac{H\beta}{H\gamma}$  and the EW of optical Fe II. Black squares: the objects with 
$\mathrm{log([O III]/H\beta}_{NLR}\mathrm{)<0.5}$, x-marks: the objects with $\mathrm{log([O III]/H\beta}_{NLR}\mathrm{)>0.5}$. }
\label{13b}
\end{figure*}

It is found that the ratio of the broad Balmer lines may be an indicator of the
 intrinsic dust extinction \citep{do2008}. However, this should not 
 strongly affect the ratio $\mathrm{H\beta}_{broad}/\mathrm{H\gamma}_{broad}$ because the extinction effects on the lines are similar (close 
 transition wavelengths).  Under some circumstances this ratio can be used for the diagnostic of the physical parameters in the BLR plasma \citep[see, e.g.][]{poo2003, il2012}.
We found that the $\mathrm{H\beta}_{broad}/\mathrm{H\gamma}_{broad}$ ratio increases as: 
(i) the FWHM H$\beta$, FWHM \ion{Mg}{2} and the Fe II$_{opt}$ width increase (see Table \ref{tbl-8});
(ii) the EW Fe II$_{opt}$ decreases (r= - 0.49, P=0, Fig \ref{13b}) and EW [\ion{O}{3}] increases (r= 0.34, P=1.3E-9).
The correlation with line widths may indicate some connections between the kinematics and physics of the emitting gas. 
 Fig \ref{13b} shows that objects with strong starburst activity (black squares) have large values of the \ion{Fe}{2} optical,
and the smallest values of the  $\mathrm{H\beta}_{broad}/\mathrm{H\gamma}_{broad}$ ratio.

Some of these ratios correlate between each other. For example, as $\mathrm{[O III]}/\mathrm{H\beta_{NLR}}$ increases,
$\mathrm{H\beta}_{broad}/\mathrm{H\gamma}_{broad}$ increases as well, but $\mathrm{Fe II}_{opt}/\mathrm{Fe II}_{UV 60}$ and
$\mathrm{Fe II}_{opt}/\mathrm{Mg II}$ decrease. The ratio $\mathrm{H\beta}_{broad}/\mathrm{H\gamma}_{broad}$ anticorrelates  
with $\mathrm{Fe II}_{opt}/\mathrm{Mg II}$ and  $\mathrm{Fe II}_{UV}/\mathrm{Mg II}$, and $\mathrm{Mg II}/\mathrm{H\beta_{broad}}$
anticorrelates with $\mathrm{Fe II}_{opt}/\mathrm{Fe II}_{UV 60}$.

Note that  the width of the narrow Balmer line 
component and the width of the broad Balmer component have opposite trends with the some ratios, especially with $\mathrm{[O III]}/\mathrm{H\beta_{NLR}}$ (see Table \ref{tbl-8}). 
  Also, the EWs of the optical \ion{Fe}{2} and [\ion{O}{3}] have opposite 
correlations with all considered ratios ($\mathrm{Fe II}_{opt}/\mathrm{Fe II}_{UV 60}$,
$\mathrm{Fe II}_{opt}/\mathrm{Mg II}$, $\mathrm{Fe II}_{UV}/\mathrm{Mg II}$,
$\mathrm{H\beta}_{broad}/\mathrm{H\gamma}_{broad}$ and $\mathrm{O III}/\mathrm{H\beta_{NLR}}$), except
with $\mathrm{Mg II}/\mathrm{H\beta_{broad}}$. This reflects the EW \ion{Fe}{2} vs. EW [\ion{O}{3}] anticorrelation. 

The averaged flux ratios of emission lines are given in Table \ref{tbl-nova2}. As it can be seen,
the optical \ion{Fe}{2} lines (in range 4400-5500 \AA) are, on average, about five times stronger than the UV ones (in range 2900-2980 \AA, multiplet 60), whereas in the SB dominant subsample
they are $\sim$7.5 times stronger than the UV ones. It is interesting that the ratio of $\mathrm{Fe II}_{opt}/\mathrm{Mg II}$ (which is expected to be a cosmological indicator) is significantly higher in the SB dominant subsample 
($\sim$1.4) compared with the AGN dominant subsample ($\sim$0.8).

\section{Discussion}

\subsection{Location of the UV \ion{Fe}{2} emitting region}

In our previous work, we discussed the location of the optical \ion{Fe}{2} emission region
\citep[see][]{po2009, ko2010, sh2012}, and found that it is located in an outer part of
the BLR (so-called the ILR), which is recently confirmed by \ion{Fe}{2} reverberation \citep[see][]{ba13}.

The correlation between the widths of the optical \ion{Fe}{2} and  UV \ion{Fe}{2} lines
 indicates that the UV \ion{Fe}{2} emission region is probably located close to the \ion{Fe}{2} optical one.
Moreover, the widths and the shifts of the UV and optical \ion{Fe}{2} lines are correlated with the widths and shifts of the \ion{Mg}{2} core and H$\beta$ ILR component (see Table \ref{tbl-2} and Table \ref{tbl-3}).  The 
average widths of the UV and optical \ion{Fe}{2} lines are very similar (2530 km s$^{-1}$ and 2360 km s$^{-1}$), and they 
are both  broader than the 
average widths of the H$\beta$ and \ion{Mg}{2} cores (1930 km s$^{-1}$ and 1590 km s$^{-1}$). 
The correlation of the iron lines width is even more significant with the FWHM of the total broad profile (wings + core) of
\ion{Mg}{2} and H$\beta$, but there is 
no a positive correlation with the separated wing component of these lines. It seems that the iron lines (both,
UV and optical) generally originate in the outer part of the BLR (ILR).
 However, as we mentioned in \cite{ko2010}, there may be an additional emission of the \ion{Fe}{2} lines 
(in the UV and optical)  that is coming from the inner part of BLR (VBLR). This emission is probably contributing to 
the continuum because the lines are very broad and cannot be resolved in the bulk of \ion{Fe}{2} lines in the
UV and optical spectral ranges. 

 In order to test the location of the forming region of the UV and optical \ion{Fe}{2} lines, we searched for the same geometry as in the \ion{Mg}{2} and H$\beta$ emission regions. 
We modified the UV and optical iron templates, assuming that the profiles of the iron lines are the same as the profiles of the \ion{Mg}{2} or H$\beta$. We compared the accuracy of
 the new fits with the previous, single Gaussian model (see Appendix A). We found that: (a) Single Gaussian profile gives better fit for the UV iron lines, compared with 
\ion{Mg}{2} and H$\beta$ profiles, (b) H$\beta$ profile fits slightly better optical \ion{Fe}{2} than single Gaussian profile. This indicates
 that, at least in the optical \ion{Fe}{2} lines, there is a contribution of the VBLR emission.

The  averaged values of line widths are similar for the UV and optical Fe II. However, there is a great difference in the the averaged values of their shifts. For the optical \ion{Fe}{2} the shift is 350$\pm$510 km s$^{-1}$, whereas
for the UV \ion{Fe}{2} is much larger: 1150$\pm$580 km s$^{-1}$. Therefore, the systemic redshift (relative to the [O III]) could be significantly detected only in the UV Fe II, but it is not significant in the optical Fe II.

The absence of the significant redshift found for the optical \ion{Fe}{2} is in the agreement with the previously obtained result: 100$\pm$240 km s$^{-1}$ \citep{ko2010}. \cite{su12}
could not confirm redshift in the optical \ion{Fe}{2} lines, as well, but \cite{hu2008a} found a slight redshift  in the optical Fe II: 407$\pm$200 km s$^{-1}$.
Note that between the shifts of the optical and UV \ion{Fe}{2} there is no a significant correlation, but only a weak trend.

We should note here that the large averaged redshift of the UV \ion{Fe}{2}  should be taken with the caution, because of uncertainties in the fit of the UV \ion{Fe}{2} in some spectra. The UV \ion{Fe}{2} lines can be very broad, which makes it difficult to resolve them from the \ion{Mg}{2} wings.

The systemic redshift probably represents the infall of the emitting gas \citep{hu2008a, fe2009}. 
\cite{hu2008a} found the correlation between the optical \ion{Fe}{2} systemic redshift and L$_{bol}/$L$_{Edd}$, and they  speculate that the inf\-low is driven by gravity
toward the center and decelerated by the radiation pressure. \cite{fe2009} investigated the geometry of the \ion{Fe}{2} emission region, taking into account the distribution of emitting clouds.
If the distribution of emitting clouds is symmetric, we see
the same number of clouds from their illuminated as from their shielded faces. But if distribution of emitters is asymmetric, we mainly observe \ion{Fe}{2} emission from the shielded face of infalling clouds, 
 which is reflected in the systemic redshift of the \ion{Fe}{2} lines \citep{fe2009}. Their calculation show that the distribution of the UV \ion{Fe}{2} emission region is more asymmetric than the distribution of the optical
Fe II. This model reproduces well the observed $\mathrm{Fe II}_{opt}/\mathrm{Fe II}_{UV}$  flux ratio.

The results from this paper, the significant average systemic redshift for the UV \ion{Fe}{2}, and absence of the significant redshift for the optical \ion{Fe}{2}, support the model given by \cite{fe2009}.
Although the UV and optical \ion{Fe}{2} emission is emitted from approximately  the same region in the AGN structure, it might be that the distribution of emitting clouds is different: the UV \ion{Fe}{2} emission clouds are probably distributed 
 asymmetrically, while the optical \ion{Fe}{2} ones have the isotropic distribution. 

 Other explanations of the results, except asymmetry of the emission region, are also possible. One can speculate that an inflow can be associated with internal shock waves, which may contribute to more
 effective excitation of the UV lines \citep[see][]{dop2005}. Namely, it is possible that UV \ion{Fe}{2} lines are more excited in infalling gas, compared with the optical \ion{Fe}{2}. For example, some connection between \ion{Mg}{2} flux and jet emission is 
 detected in \cite{lt2013}.

\subsection{Peculiarities of the optical \ion{Fe}{2} correlations }

Concerning correlations between the optical \ion{Fe}{2} lines and analyzed UV/optical lines, we can 
point out several peculiarities:
\begin{enumerate}
\item Only the EW of
the optical \ion{Fe}{2} anticorrelates
with the EW of some other line (EW [O III]), while all other lines correlate with EWs of other lines or show no correlations (Table \ref{tbl-7}). 
This specific anticorrelation is reflected through correlations shown in Table \ref{tbl-8}, as well, where the EW \ion{Fe}{2} and EW [\ion{O}{3}] 
have opposite correlations with considered ratios.  
\item 

The  EWs of analyzed lines (UV \ion{Fe}{2}, H$\beta$ NLR and \ion{Mg}{2}) increase as their widths increase, or there is no any correlation between their 
EWs and widths as for H$\beta_{broad}$ and [\ion{O}{3}] (Table \ref{tbl-9}).
Only in the case of the optical Fe II, their EW increases as the width decreases, i.e. the \ion{Fe}{2} lines are stronger as they are narrower.
In addition, while EWs of the broad lines do not depend on widths of the other lines, the EW of the optical \ion{Fe}{2} becomes stronger as other broad 
lines are narrower (Mg II, H$\beta$). It is interesting that it is the same with the EW H$\beta$ NLR, which increases as the width of the broad lines decreases.
\end{enumerate}
On the other hand, the UV \ion{Fe}{2} lines do not show any of these peculiar correlations as the optical Fe II, e.i. it seems that their 
emission properties are quite different than for 
the optical Fe II.

\subsection{Differences between the UV and optical \ion{Fe}{2} lines}

As it can be seen in previous sections, there are significant differences between the optical and UV \ion{Fe}{2} lines, as reflected in 
different correlations of these lines with the other spectral properties. Between the EWs of these lines there is no correlation (see Sec 3.2), 
and their flux ratio depends on different physical parameters, such as column density and microturbulence 
\citep[see][]{jo1987,ve2003,ve2004,sa2011}. These differences could be explained in two ways: the emission regions of optical and UV 
\ion{Fe}{2} lines have a different spacial distribution, or the mechanisms of their excitation are not the same. 
The mixture of these two influences is also possible.

\cite{jo1987} suggested as one of the solutions that the optical and UV \ion{Fe}{2} could be emitted in two distinct regions. 
\cite{fe2009} concluded that iron lines are emitted by clouds that are distributed asymmetrically: the UV \ion{Fe}{2} lines are beamed 
toward a central source while the optical \ion{Fe}{2} lines are emitted isotropically. In the case of asymmetrical distribution, 
photoionization models can reproduce the observed UV to optical \ion{Fe}{2} flux ratio \citep[][]{sa2011}.  Our analysis of the 
kinematical properties of the optical and UV \ion{Fe}{2} lines indicates that their emission regions are located close to one another in the AGN 
structure. Also, unlike optical \ion{Fe}{2} emission region, the UV \ion{Fe}{2} emission region is probably asymmetric. Different distributions of the
 iron emission clouds in an emission region can explain their observed flux ratios, but it still cannot explain the peculiar correlations of the optical \ion{Fe}{2} 
lines with some line properties, which is not detected for the UV \ion{Fe}{2}.

 It is possible that UV and optical \ion{Fe}{2} arise with the domination of different excitation mechanisms, which could explain their 
 differences. \cite{jo1987} found  that strong optical \ion{Fe}{2} emitters can be explained with the collisional excitation in a dense 
 and cold medium 
(6000 K $<$ T $<$ 8000 K, n$_H>10^{11}$ cm$^{-3}$), while \ion{Fe}{2} UV intensities are more difficult to account for. 
The models of \cite{jo1987} show that optical and UV \ion{Fe}{2} have different correlations with the column density, due the larger optical depth of the UV \ion{Fe}{2} lines. 
For N$_H >$ 21 cm$^{-2}$, the optical \ion{Fe}{2} lines increase more rapidly than the UV \ion{Fe}{2} ones, because of a smaller optical thickness. 
For a larger column density (N$_H >$ 22 cm$^{-2}$), the UV \ion{Fe}{2} flux decreases more rapidly than the flux of the optical 
\ion{Fe}{2} \citep[see][]{jo1987}. It is the reason why the flux ratio of $\mathrm{Fe II}_{opt}/\mathrm{Fe II}_{UV}$ increases with increasing
of the column density.

\cite{sa2011} found that $\mathrm{Fe II}_{opt}/\mathrm{Fe II}_{UV}$ ratio increases with increasing of the Eddington ratio. 
A high Eddington ratio is related to dense medium and large column density, 
because the clouds with a low density and small column density would be blown away by a large radiative pressure
\citep{do2009,do2011}.  

 On the other hand, the large column density and  the dense 
environment may be also due to an increase of the star formation rate \citep[][]{ne2004,ha2013,cl2013}.  \cite{ne2004} found that a violent starforming activity can produce high-density and large column density gas in nuclear regions. 

The presence of starforming/starburst regions could be related with AGN evolution \citep{m09,sa10}. 
The spectral properties of AGNs are probably changing during the time,
and it is expected that young AGNs have a higher Eddington ratio, higher star formation rate, and smaller black hole mass and FWHMs of the broad lines 
 \citep[][]{lt06,sa10}. 

 Since  the objects in this sample have a redshift in range 0.407$<$z$<$0.643, we cannot explore evolution at high redshift, but we cannot exclude that objects with similar z could be
 in different evolution phase.
This may explain some correlations that are typical for the optical \ion{Fe}{2} lines 
 found in this paper (see Sec. 4.2), as well as correlations which are part 
of Boroson and Green  Eigenvectror 1 \citep[][]{bg92}: EW \ion{Fe}{2} optical vs. EW [\ion{O}{3}] and  EW \ion{Fe}{2} optical vs. FWHM H$\beta$. 
The anticorrelation 
of the optical \ion{Fe}{2} and [\ion{O}{3}] equivalent widths could be caused by an increase in  density and the column density  due to the influence of the starbursts: 
as the column density increases, the flux 
of the optical \ion{Fe}{2} increases, but the flux of the forbidden [\ion{O}{3}] lines decreases, because of the collisional suppression  or the weak ionizing continuum from starbursts.  
This anticorrelation is not seen for EWs of the UV \ion{Fe}{2} and [\ion{O}{3}], because the UV \ion{Fe}{2} lines decrease more rapidly with increasing  column density, compared with the optical 
\ion{Fe}{2} \citep{jo1987}. 
\cite{sa2011} suggested that increasing column density, caused by a large Eddington ratio, is a physical cause behind 
the Boroson and Green  Eigenvectror 1 correlations, but they explain the decrease of the [\ion{O}{3}] lines as inability of 
ionizing photons emitted from the central object to reach the NLR clouds, because the large-column-density clouds in the BLR. In this case, 
we would expect the anticorrelation between the EWs of the optical \ion{Fe}{2} and other narrow emission lines (e.g. narrow component of H$\beta$),
but these correlations are not observed \citep[see][]{ko2011}. The pure recombination lines, as the NLR H$\beta$, are not influenced by higher densities, while forbidden [\ion{O}{3}] lines
 would be weaker in a dense medium because of the collisional suppression.  Also, unlike the [\ion{O}{3}] lines, the Balmer lines are strongly ionized by starburst continuum.

 \cite{sa10} investigated a sample of the AGNs from the local Universe (z $<$ 0.2) using the Spitzer data and found that star formation rate is higher in objects with low FWHMs of the broad lines,
 low black hole mass and high Eddington ratio.
Also, in model of \citet{lt06} young AGNs are expected to have the FWHM of the broad lines smaller than the older ones, because the BLR is not
developed yet. This may explain other correlations typical for the optical Fe II, as e.g. an anticorrelation
EW \ion{Fe}{2} optical vs. FWHM H$\beta$. The \ion{Fe}{2} optical emission is stronger in a high density/large column density environment,
which is expected in young objects with the high Eddington ratio, high star formation rate and relatively narrower broad emission lines.
In our sample, we also find an anticorrelation between the EW \ion{Fe}{2} optical and the widths of the \ion{Fe}{2} optical and Mg II.

 If we assume that the ratio $\mathrm{[O III]}/\mathrm{H\beta_{NLR}}$ is an indicator of the starburst fraction (the increasing ratio 
reflects the decreasing starburst influence), than the 
 found correlations of this ratio with the other spectral properties supports this model. In Table \ref{tbl-8} and Figs 
 \ref{6aaa}, \ref{13aa}, \ref{14}, \ref{13}, \ref{14a} and \ref{13b} it could be seen that for the strong starbust influence
(small values of the $\mathrm{[O III]}/\mathrm{H\beta_{NLR}}$ ratio), the broad lines are narrower, the narrow lines are broader, 
EW of the optical \ion{Fe}{2} is stronger and the EW of the [\ion{O}{3}] is weaker. In difference
with the optical Fe II, the EW of UV \ion{Fe}{2} does not depend on the starburst influence (see Table \ref{tbl-8} and
Fig \ref{13aa}). Also the anticorrelation of this ratio with the ratios $\mathrm{Fe II}_{opt}/\mathrm{Fe II}_{UV 60}$ and 
$\mathrm{Fe II}_{opt}/\mathrm{Mg II}$,
which depend on the column density and microturbulence, imply that for starburst dominant objects, the column 
density is larger and microturbulence is smaller, compared with the gas in pure AGNs.

A high density may stimulate some excitation processes that have a more important role for the optical Fe II. 
This 
could be the collisional excitation or increased Ly$\alpha$ fluorescence,  which may be one of the very important 
additional mechanisms of 
the \ion{Fe}{2} excitation \citep[][]{pe1987,gr1996,sp1998,ba2004}.

 The model we have is discussed here certainly is not an unique interpretation of the data. We may speculate about some other explanations of these correlations. For example, they could be consequence of the different viewing angle. It 
is possible that visibility of the optical \ion{Fe}{2} emission region increases as we get a more pole-on view of the BLR, while the broad line widths decrease. 
This could be reflected as an anticorrelation of the optical \ion{Fe}{2} strength and broad line width. There are some indications that EW [\ion{O}{3}] also depends on 
inclination \citep[see][]{ri2011}, which can be the explanation for the anticorrelation of the optical \ion{Fe}{2} lines and [\ion{O}{3}]. In this case, these anticorrelations
are not seen for the UV \ion{Fe}{2} lines because their emission region has the different spacial distribution than the optical \ion{Fe}{2} one.


\section{Conclusions}

In this paper we investigate the connections between the UV and optical \ion{Fe}{2} emission lines 
using a sample of the 293 AGNs from the SDSS database. The properties of the optical and UV iron lines are compared and correlated
with the properties of the other  emission lines that are present in the observed spectral range, in order to investigate the origin of the iron lines 
 and processes responsible for their emission.
We model the UV \ion{Fe}{2} emission taking into account the contribution of the different multiplets \citep[][]{po2003}. Additionally, 
we use a new model for the Balmer continuum subtraction \citep[][]{ko2014}. The strong 
emission lines are decomposed into components that are coming from the different emission regions,  in order to explore and compare the physical properties of the environment where the components arise. The flux ratios of the  lines, which can be the indicators of the physical 
conditions in the emission region, are also analyzed. 
We consider the influence of  starburst activity to the spectral properties of the iron emission.
After investigation of correlations between the UV/optical \ion{Fe}{2} and other lines from the spectral range,
we can outline the following conclusions: 

\begin{enumerate}

\item The UV and optical \ion{Fe}{2} lines arise in the close emission regions in the AGN structure. Most of the \ion{Fe}{2} UV/optical
emission is probably  originating in the Intermediate Line emission Region i.e. a region with Doppler velocities 
around 1500-2000 $\rm km s^{-1}$. However, the \ion{Fe}{2} lines tend to be broader than the ILR components of
H$\beta$ and \ion{Mg}{2}, which indicates a small contribution  of VBLR emission, at least in the \ion{Fe}{2} optical lines.

\item  The significant systemic redshift, which is the signature of the gas infall, is found only for the UV \ion{Fe}{2} lines (1150$\pm$580 km s$^{-1}$), but not for the optical \ion{Fe}{2} (350$\pm$510 km s$^{-1}$). 
 This indicates that, although the UV and optical \ion{Fe}{2} emission clouds
are located in approximately the same region of the AGN structure, their
distribution is probably different: the UV \ion{Fe}{2} emission clouds seem to be distributed asymmetrically (we see more shielded, infalling clouds)
while the optical \ion{Fe}{2} emission clouds are probably isotropically distributed.  Except emission region asymmetry, some other models can explain high UV \ion{Fe}{2} redshift, e.g. more efficient excitation of the UV \ion{Fe}{2} lines in the infalling gas, due to shock waves.

\item  There are significant differences between the optical and UV \ion{Fe}{2} lines, which are presented in 
different correlations between these lines and other spectral properties. The intriguing anticorrelations found for the optical \ion{Fe}{2} (EW Fe II$_{opt}$ vs. EW [\ion{O}{3}] and EW Fe II$_{opt}$ vs. FWHM H$\beta$), which are part of the \citet[][]{bg92} Eigenvector 1,
are not detected for the UV Fe II. Beside all analyzed broad lines, only the EW Fe II$_{opt}$ anticorrelates with EW some other emission line ([\ion{O}{3}]), 
and only EW Fe II$_{opt}$ increases as the widths of all broad lines (including Fe II$_{opt}$) decrease.

\item The peculiar anticorrelations of the optical \ion{Fe}{2} are
probably connected with a  high density and a large column density region in AGNs with a high star formation rate.
The anticorrelation between EWs of the optical \ion{Fe}{2} and [\ion{O}{3}] lines is probably due increase of the additional excitation mechanism of the 
optical \ion{Fe}{2} with the increase of the column density, and at the same time, a decrease of the forbidden [\ion{O}{3}] lines due collisional suppression, or because of low ionization continuum from starbursts. 
  On the other hand, the anticorrelation EW Fe II$_{opt}$ vs. FWHM H$\beta$ is present because the AGNs, with high star formation rate and large column density regions, are expected to be young objects with a smaller mass of the black hole and widths of the broad lines that are generally narrower
 than in the spectra of older, pure AGNs.  These correlations are not detected in the UV \ion{Fe}{2} lines, since their optical thickness is larger than for optical \ion{Fe}{2}, so with increasing column density, UV \ion{Fe}{2} lines decrease more rapidly than optical \ion{Fe}{2}. 
 Other explanations of the these correlations are possible. For example, they could be caused by different angle of view.
\end{enumerate}

Additionally, we explore some correlations between the \ion{Mg}{2} and Balmer lines, as well as between the flux ratio of different lines and 
kinematical parameters. From this investigation we can outline following conclusions:
\begin{enumerate}
\item There is an expected good correlation between FWHMs of the broad \ion{Mg}{2} and H$\beta$ line. Both, the
\ion{Mg}{2} and broad Balmer lines can be decomposed into two components, the VBLR and ILR, which shows a complex structure of the BLR. Moreover, there is an
anticorrelation between the width and shift of the \ion{Mg}{2} wing component, that indicates 
the asymmetry in the Mg II, i.e. the narrower
\ion{Mg}{2} wing component tends to be asymmetric. 
\item It seems that the explored line ratios (see Table \ref{tbl-8}) are connected with the FWHMs of broad lines, especially with FWHM Mg II, where 
a positive trend between FWHM \ion{Mg}{2} and ratios $\mathrm{H\beta}_{broad}/\mathrm{H\gamma}_{broad}$ and $\mathrm{Mg II}/\mathrm{H\beta_{broad}}$  is found, and anticorrelations with 
 $\mathrm{Fe II}_{opt}/\mathrm{Fe II}_{UV 60}$ and $\mathrm{Fe II}_{opt}/\mathrm{Mg II}$. This is an indicator that the physical processes and 
abundances in the BLR, which are reflected in the ratios, are connected with the emission region kinematics.
\end{enumerate}

\acknowledgments

This work is a part of the project (146002) ``Astrophysical
Spectroscopy of Extragalactic Objects'' supported by the Ministry of
Science of Serbia. The authors would like to thank the Alexander Von Humboldt (AvH) foundation for its support this
work through project "Probing the Structure and Physics of the BLR using AGN Variability" in the frame of the
AvH program for funding a research group linkage.

\appendix
\section{The fitting of the iron lines with Mg II and H$\beta$ profiles}

\begin{figure*}
\includegraphics[width=0.40\textwidth]{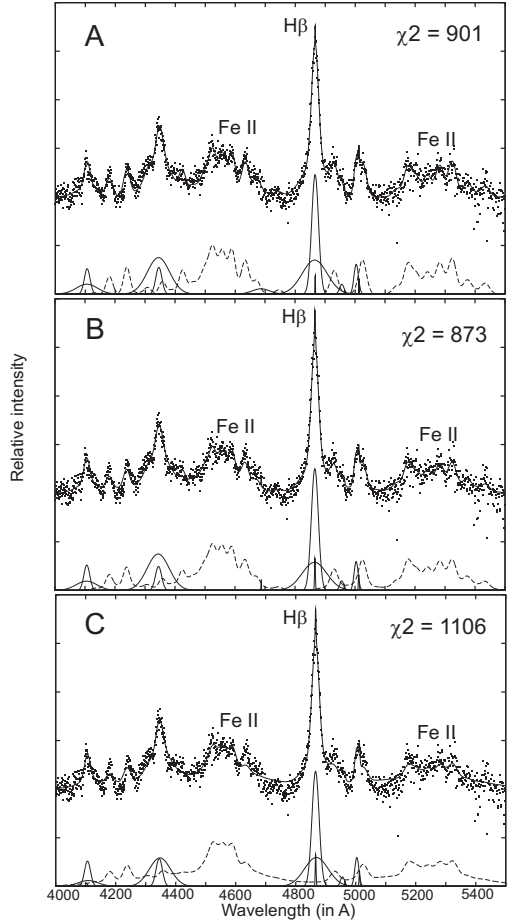}

\caption{Example of the fit of the optical iron lines with single Gaussian model (A), with H$\beta$ profile (B) and Mg II profile (C), for object: SDSS J020435.18$-$093154.9. The iron template is denoted with dashed line.}
\label{15}
\end{figure*}

\begin{figure*}

\includegraphics[width=0.35\textwidth]{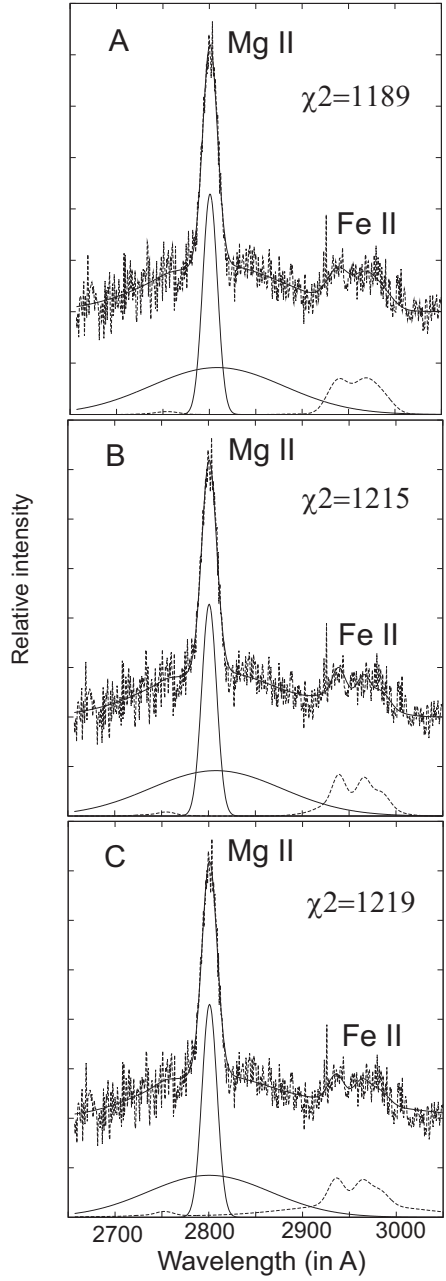}
\caption{Same as in previous figure, but for the UV iron lines.}
\label{16}
\end{figure*}

 In order to fit the optical iron lines with the H$\beta$ line profile, we modify the iron template, so that instead of one Gaussian for each iron line, there are two Gaussians with the same widths, relative shifts, and relative intensities as the ILR and VBLR H$\beta$ components. We keep the same
 relative intensities between the iron lines in the multiplets  as in the initial, single Gaussian template, and the shift of the template is the free parameter. We fit simultaneously the H$\beta$ 
line and the optical \ion{Fe}{2} template made of the sum of H$\beta$ profiles. 
After that, we applied the other template made of the \ion{Mg}{2} profiles, where the widths, relative shifts, and
relative intensities of the \ion{Mg}{2} core and the wings are fixed values obtained from the fit in the UV part of spectrum. We repeat this procedure for the UV \ion{Fe}{2}
 lines using \ion{Mg}{2} and H$\beta$ profiles in the UV \ion{Fe}{2} template in the same way, but this time \ion{Mg}{2} line is fitted simultaneously with UV Fe II template, which is made of \ion{Mg}{2} 
profiles. For template with  H$\beta$ profiles we use the fixed values of H$\beta$ parameters, which were obtained from the fit in the optical range.
An example of the one object fit (SDSS J020435.18$-$093154.9) with the different iron templates is given in Figs \ref{15} and \ref{16}. It can be seen that fit of
 the optical \ion{Fe}{2} lines with H$\beta$ profiles (B) is slightly better
than with single Gaussian model (A), whereas the iron template with \ion{Mg}{2} profiles cannot fit well the optical iron lines (C). In the case of the UV \ion{Fe}{2} lines, the single Gaussian 
model (A) fits the iron lines better, than other two templates (B and C).

We compare the $\chi^2$ of the fits obtained from the different templates with the $\chi^2$ of the initial fit where the iron lines are fitted with one Gaussian
 for each line. The results for the whole sample are given in Table \ref{App}. Generally, the differences between $\chi^2$ obtained with different templates are, for most of objects, small and less than 5\%.
The fit of the optical iron lines with the H$\beta$ profiles gives a slightly better fit, comparing the single Gaussian model for $\sim$85 \% of objects from the sample. 
Only in 3\%  of objects, fit is significantly improved (the difference between $\chi^2$ higher than 10\%).
In the case of the fit of optical iron lines with the \ion{Mg}{2} profile, only in the half of the sample fits are better. The fit of the UV iron lines is less accurate with 
H$\beta$ and \ion{Mg}{2} profiles than with one Gaussian model for the majority of objects from the sample.

\clearpage

\clearpage


\clearpage

\begin{table}
\begin{center}

\caption{The list of the strongest UV \ion{Fe}{2} lines within $\lambda\lambda$ 2650-3050 \AA \ range used in model. 
 \label{tbl-1}}
\begin{tabular}{c c c c}
\tableline\tableline

Wavelength&Multiplet&Transitions&Relative intensity\\
\tableline\
\vspace{-2mm}
2926.58&60 & a${\ }^4D_{7/2}$ - z${\ }^6F^o_{9/2}$& 1.000\\
\vspace*{-2mm}
2953.77&60 & a${\ }^4D_{5/2}$ - z${\ }^6F^o_{7/2}$ & 0.842\\
\vspace*{-2mm}
2970.51 &60 & a${\ }^4D_{3/2}$ - z${\ }^6F^o_{5/2}$ &0.386\\
\vspace*{-2mm}
2979.35 &60 &  a${\ }^4D_{1/2}$ - z${\ }^6F^o_{3/2}$& 0.165\\
\vspace*{-2mm}
2916.15 &60 &  a${\ }^4D_{7/2}$ - z${\ }^6F^o_{7/2}$&0.007\\
\vspace*{-2mm}
2945.26 &60 &  a${\ }^4D_{5/2}$ - z${\ }^6F^o_{5/2}$&0.007\\
\vspace*{-2mm}
2975.94 &60 & a${\ }^4D_{1/2}$ - z${\ }^6F^o_{1/2}$ &  0.039\\
\vspace*{-2mm}
2907.85 &60 & a${\ }^4D_{7/2}$ - z${\ }^6F^o_{5/2}$ &0.015\\
\vspace*{-2mm}
2939.51 &60 & a${\ }^4D_{5/2}$ - z${\ }^6F^o_{3/2}$ &0.030\\
2961.27&60 & a${\ }^4D_{3/2}$ - z${\ }^6F^o_{1/2}$ & 0.035\\

\tableline
\vspace*{-2mm}
2880.75 &61 &a${\ }^4D_{7/2}$ - z${\ }^6P^o_{7/2}$  & 1.000\\
\vspace*{-2mm}
2868.87 &61 &a${\ }^4D_{5/2}$ - z${\ }^6P^o_{5/2}$ & 0.229\\
\vspace*{-2mm}
2861.19 &61 &a${\ }^4D_{3/2}$ - z${\ }^6P^o_{3/2}$ & 0.042\\
\vspace*{-2mm}
2917.46 &61 &a${\ }^4D_{5/2}$ - z${\ }^6P^o_{7/2}$ & 0.071\\
2892.82 &61 &a${\ }^4D_{3/2}$ - z${\ }^6P^o_{5/2}$ & 0.062\\

\tableline
\vspace*{-2mm}
2755.73&62 &a${\ }^4D_{7/2}$ - z${\ }^4F^o_{9/2}$ & 1.000\\
\vspace*{-2mm}
2749.33 &62 &a${\ }^4D_{5/2}$ - z${\ }^4F^o_{7/2}$ & 0.773\\
\vspace*{-2mm}
2749.74! &62 &a${\ }^4D_{5/2}$ - z${\ }^4F^o_{3/2}$ &0.104\\
\vspace*{-2mm}
2746.48 &62 &a${\ }^4D_{3/2}$ - z${\ }^4F^o_{5/2}$ & 0.545\\
\vspace*{-2mm}
2743.20&62 &a${\ }^4D_{1/2}$ - z${\ }^4F^o_{3/2}$ & 0.332\\
\vspace*{-2mm}
2716.68&62 &a${\ }^4D_{7/2}$ - z${\ }^4F^o_{7/2}$ & 0.0003\\
\vspace*{-2mm}
2724.88 &62 &a${\ }^4D_{5/2}$ - z${\ }^4F^o_{5/2}$ & 0.264\\
\vspace*{-2mm}
2730.73&62 &a${\ }^4D_{3/2}$ - z${\ }^4F^o_{3/2}$ & 0.045\\
2709.37 &62 &a${\ }^4D_{5/2}$ - z${\ }^4F^o_{3/2}$ & 0.0004\\
\tableline
\vspace*{-2mm}
2739.54 &63 &a${\ }^4D_{7/2}$ - z${\ }^4D^o_{7/2}$ &1.000\\
\vspace*{-2mm}
2746.98 &63 &a${\ }^4D_{5/2}$ - z${\ }^4D^o_{5/2}$ & 0.653\\
\vspace*{-2mm}
2749.18 &63 &a${\ }^4D_{3/2}$ - z${\ }^4D^o_{3/2}$ &0.300\\
\vspace*{-2mm}
2749.48 &63 &a${\ }^4D_{1/2}$ - z${\ }^4D^o_{1/2}$ &0.153\\
\vspace*{-2mm}
2714.41 &63 & a${\ }^4D_{7/2}$ - z${\ }^4D^o_{5/2}$&0.220\\
\vspace*{-2mm}
2727.54&63 &a${\ }^4D_{5/2}$ - z${\ }^4D^o_{3/2}$ &0.227\\
\vspace*{-2mm}
2772.72 &63 &a${\ }^4D_{5/2}$ - z${\ }^4D^o_{7/2}$ & 0.0003\\
\vspace*{-2mm}
2768.94  &63 &a${\ }^4D_{3/2}$ - z${\ }^4D^o_{5/2}$ & 0.019\\
2761.81 &63 &a${\ }^4D_{1/2}$ - z${\ }^4D^o_{3/2}$ & 0.031\\

\tableline

\end{tabular}
\end{center}
\end{table}

\clearpage

\begin{deluxetable}{ccccccccccc}
\rotate
\tabletypesize{\tiny}
\tablewidth{0pt}
\tablecaption{\scriptsize The correlations between widths (w), and between widths (w) and shifts (sh) of the optical and UV lines. 
The Balmer line compoments are denoted as: NLR, ILR and VBLR. All shifts are measured relative to the [\ion{O}{3}] 5007  \AA. The Spearman coefficient of correlation (r) and P-value 
are given in bold print for correlations with P$<$1E-9. \label{tbl-2}}
\tablehead{

\colhead{} & \colhead{} & \colhead{w NLR} & \colhead{w ILR} &\colhead{w VBLR}& \colhead{FWHM H$\beta$}&\colhead{w Mg 
II core} &\colhead{w Mg II wings}&\colhead{FWHM Mg II}&\colhead{w Fe II$_{opt}$}&\colhead{w Fe II$_{UV}$}
 }
\startdata
w NLR&r & 1&-0.03&-0.03&-0.1  &-0.19 &0.11&-0.19&-0.25&-0.10  \\
&P & 0 & 0.59 & 0.56& 0.09 &1.3E-3&0.07&8.6E-4&2E-5& 0.08 \\
\tableline
w ILR&r &-0.03&1&{\bf 0.35}&{\bf 0.83}&{\bf 0.60}&-0.28&{\bf 0.63}&{\bf 0.58}&{\bf 0.39} \\
&P &0.59& 0&{\bf 9.8E-10}&{\bf 0}&{\bf 0}&1.4E-6&{\bf 0 }&{\bf 0 }&{\bf 2.4E-12} \\
\tableline
w VBLR&r &-0.03&{\bf 0.35}&1 &0.30 &0.21 & -0.15&0.27&0.19&0.20 \\
&P &0.56&{\bf 9.8E-10}&0&1.3E-7  &2.5E-4&8.6E-3&2.8E-6&9.3E-4&7.0E-4\\
\tableline
w Mg II core&r &-0.19 &{\bf 0.60} & 0.21&{\bf 0.67 } &1&-0.3& {\bf 0.85}& {\bf 0.57}& {\bf 0.49}  \\
&P &1.3E-3&{\bf 0}&2.5E-4&{\bf 0}& 0 &2.2E-7&{\bf 0}& {\bf 0}&{\bf 0}  \\
\tableline
w Mg II wings&r &0.11&-0.28 &-0.15&{\bf -0.36 } &-0.30 &1&-0.35&{\bf -0.37}&{\bf -0.43} \\
&P & 0.07&1.4E-6& 8.6E-3& {\bf  2.7E-10}&2.2E-7&0&9.3E-10&{\bf 5.4E-11 }& {\bf 6.4E-15 }\\
\tableline
w Fe II$_{opt}$&r & 0.25&{\bf 0.58}& 0.19&{\bf 0.69} &{\bf 0.57}& {\bf -0.37}&{\bf 0.64}&1&{\bf 0.39}  \\
&P &2.0E-5&{\bf 0}&9.3E-4&{\bf 0 } &{\bf 0}&{\bf 5.4E-11}&{\bf 0}&0&{\bf 4.1E-12} \\
\tableline
w Fe II$_{UV}$&r  &-0.10&{\bf 0.39} &0.20&{\bf 0.47}&{\bf 0.49}&{\bf -0.43}&{\bf 0.48}&{\bf 0.39}&1\\
&P &0.08&{\bf 2.4E-12}&7.0E-4&{\bf  0}&{\bf  0} &{\bf 6.4E-15}&{\bf  0} &{\bf 4.1E-12}&0\\
\tableline

sh ILR   &r&0.13  &   -0.15  & -0.07  &   -0.26   &   -0.26   & 0.13   &   -0.26   &-0.25    & -0.18 \\
&P          & 0.03 &   0.01 &  0.21 &   9.4E-6 &  7.8E-6 & 0.03         &   6.3E-6 & 1.0E-5  & 2.0E-3\\
\tableline
sh VBLR &r   &-0.02 &{\bf 0.42}   &  0.30  &  {\bf 0.39}    &  {\bf 0.36}     &  -0.30    &   {\bf 0.38}    &  0.30         & 0.24\\
&P          & 0.73 &{\bf 3.4E-14}& 2.1E-7 & {\bf 5.9E-12}  &  {\bf 3.0E-10}  & 2.6E-7    &    {\bf 1.8E-11}& 1.1E-7        &3.3E-5\\
\tableline
sh Mg II core &r  & 0.09  & 0.02   & -0.08   &  -0.01 &  0.04 &  0.04  &  0.002  &  0.03  &  -0.08\\
&P               & 0.12  & 0.71   & 0.18    &   0.81 &   0.50  &  0.45  & 0.98   &   0.64 &   0.18\\
\tableline
sh Mg II wings &r&  -0.05  &  0.07  &  -0.03  &  0.14  &  0.17  &  {\bf -0.56}  &  0.11  &  0.13  &  {\bf 0.46}\\
&P               &0.43    &  0.25  &    0.56 & 0.02   &   4E-3 &  {\bf 0}      &  0.05 &  0.03  &{\bf 2.2E-16}\\
\tableline
 sh Fe II$_{opt}$ &r  &  0.003  &     0.21   & 0.07  &  0.24  &  0.24  &  -0.10  &  0.24  &  0.34 & 0.13\\
&P               &  0.97   &   2.4E-4   & 0.21   &  3.5E-5&  4E-5  &  0.08   &  2.4E-5& 1.7E-9  &0.03\\
\tableline
sh Fe II$_{UV}$ &r   &0.04  &    0.1     &   0.0037 &   0.06 &  -0.025  &  0.25 &  0.02  &  0.05  &  -0.1  \\
&P               &0.49     &    0.1  & 0.95   &   0.33 &  0.66   & 1.0E-5   &  0.72    &  0.37  & 0.1\\

\tableline
\enddata
\end{deluxetable}

\newpage

\begin{deluxetable}{ccccccccc}
\tabletypesize{\scriptsize }
\tablecaption{The correlations between velocity shifts (sh) of the optical and UV lines. The broad Balmer line components are denoted as:
ILR and VBLR. All shifts are measured relative to the [\ion{O}{3}] 5007  \AA. The Spearman coefficient of correlation (r) and P-value  are given in bold print for correlations with
P$<$1E-9. \label{tbl-3}}
\tablehead{

\colhead{} & \colhead{}  & \colhead{sh ILR} &\colhead{sh VBLR}& \colhead{sh Mg II
core} &\colhead{sh Mg II wings}&\colhead{sh Fe II$_{opt}$}&\colhead{sh Fe II$_{UV}$}
 }
\startdata

sh ILR&r        & 1 &-0.13 &{\bf 0.62}    & -0.05  & 0.34  & 0.32\\
&P              &  0& 0.02 & {\bf 0}      & 0.39   & 1.6E-9& 3.6E-8\\
\tableline
sh VBLR&r       &-0.13  &1  &0.20    &0.22   &0.20    & 0.12\\
&P              & 0.02  &0  & 4.5E-4 & 1.6E-4& 5.9E-4 & 0.04\\
\tableline
sh Mg II core&r & {\bf 0.62}& 0.20    & 1  &  0.02&{\bf 0.40 }    &{\bf 0.48}\\
&P              & {\bf 0}   & 4.5E-4  &  0 &  0.71&{\bf 1.05E-12} & {\bf 0}\\
\tableline
sh Mg II wings&r  &-0.05  &0.22  &0.02 & 1&0.03& -0.21\\
&P                &0.39   &1.6E-4& 0.71&0 &0.62 & 3.6E-4\\
\tableline
sh Fe II$_{opt}$ &r &0.34& 0.20    &{\bf0.40 }    &0.03& 1&0.23 \\
&P              &1.6E-9& 5.9E-4&{\bf1.1E-12 } &0.62& 0& 7.1E-5 \\
\tableline
sh Fe II$_{UV}$   &r&0.32&0.12&{\bf 0.48}&-0.21&0.23& 1  \\
&P              &3.6E-8& 0.04& {\bf 0}& 3.6E-4&7.1E-5&0\\

\tableline
\enddata
\end{deluxetable}

\clearpage


\begin{deluxetable}{ccccc}
\tablecaption{The average values and standard deviations (SD) for Doppler widths and velocity shifts of the observed 
lines and their components (given in kms$^{-1}$). The shifts are measured relative to the [\ion{O}{3}] 5007  \AA.
 \label{tbl-4}}
\tablehead{

\colhead{} & \multicolumn{2}{c} { width}  &  \multicolumn{2}{c}{velocity shift} \\
\colhead{} & \colhead{average values} & \colhead{ SD} & \colhead{average values} & \colhead{ SD} 
 }
\startdata

 Balmer line NLR  &310  & 170 &0  & 0 \\
\tableline
 Balmer line ILR  &1930  & 650 & -50 & 400\\
\tableline
Balmer line VBLR  &4370  &980 & 770 &1120 \\
\tableline
Mg II 2800 core &1590  &390 & -3  & 240 \\
\tableline
Mg II 2800 wings  & 7770 & 1790 & 500 & 1260\\
\tableline
Fe II optical  &2360  &1050  & 350 & 510\\
\tableline
Fe II UV  &2530  &990 &1150 & 580 \\

\enddata

\end{deluxetable}

\clearpage

\begin{deluxetable}{ccccccc}
\tabletypesize{\scriptsize }
\tablecaption{The average values and standard deviations (SD) for EWs of observed lines and their components,
calculated for the total sample (293 objects), and for subsamples
with $\mathrm{log([O III]/H\beta}_{NLR}\mathrm{)>0.5}$ (AGN dominant, 247 objects ) and $\mathrm{log([O III]/H\beta}_{NLR}\mathrm{)<0.5}$ (starburst - 
SB dominant, 46 objects). 
 \label{tbl-nova1}}
\tablehead{

\colhead{} & \multicolumn{2}{c} { ALL}  &  \multicolumn{2}{c}{AGN dominant}&  \multicolumn{2}{c}{SB dominant} \\
\colhead{} & \colhead{average values} & \colhead{ SD} & \colhead{average values} & \colhead{ SD} & \colhead{average values} & \colhead{ SD} 
 }
\startdata

EW Fe II UV$_{total}$   &           11.879  &         8.380& 12.093  &  8.899  &  10.731 & 4.596  \\
EW Mg II$_{wings}$      &          29.279    &        11.790& 30.457  &  12.065 &  22.956 & 7.605   \\
EW Mg II$_{core}$    &           15.389    &         8.849&   15.874&  9.404  &  12.781 & 4.044 \\
EW Fe II opt$_{total}$&         101.216      &       42.760 &   94.058 &  40.211  &  139.653& 35.123    \\
EW [O III] 5007     &           17.051       &        14.806&   18.767&  15.305 &  7.834  & 6.232    \\
EW H$\beta$ NLR        &        4.210       &     12.945  &  3.255  &  11.198 &  9.334  & 19.236 \\
EW H$\beta$ ILR        &        32.409     &      13.588  &  32.634 &  13.782 &  31.197 & 12.566    \\
EW H$\beta$ BLR        &        39.721      &      16.720  &  42.027 &  16.638&  27.343 & 10.721   \\
EW H$\gamma$ NLR        &        0.558      &     0.581   & 0.499  &  0.506   &  0.873   & 0.817   \\
EW H$\gamma$ broad      &          28.776     &       7.541 & 28.848  &  7.781  &  28.389& 6.158  \\
EWH$\delta$ NLR        &        0.272     &     0.362    &  0.239  &  0.318   &  0.452 & 0.507    \\
EWH$\delta$ broad      &          7.728      &      3.340   &  7.837   &  3.452  &  7.140  & 2.608        \\
\tableline

 \enddata
\end{deluxetable}

\clearpage

\begin{deluxetable}{cccccccccccc}
\rotate
\tabletypesize{\tiny}
\tablewidth{0pt}
\tablecaption{ The correlations between the equivalent widths (EWs) of optical and UV lines.   \label{tbl-7}}
\tablehead{
\colhead{} &
\colhead{} &
\colhead{\scriptsize FeII UV\tablenotemark{a}} &\colhead{\scriptsize FeII$_{opt}$\tablenotemark{b} } &\colhead{\scriptsize [OIII]} &\colhead{\scriptsize MgII$_{total}$\tablenotemark{c}} &\colhead{\scriptsize H$\beta_{NLR}$ }&
\colhead{\scriptsize H$\beta_{broad}$\tablenotemark{d} } &\colhead{\scriptsize H$\gamma_{NLR}$ } &\colhead{\scriptsize H$\gamma_{broad}$} &\colhead{\scriptsize H$\delta_{NLR}$ } &\colhead{\scriptsize H$\delta_{broad}$ }  }

\startdata
FeII UV & r& 1  &    0.15 &   -0.11 & {\bf 0.39} &   -0.10 &  0.005 &  -0.16&     0.17  &     -0.04 &   -0.05   \\ 
        & P& 0  &    0.01  &   0.06& {\bf 3.2E-12} & 0.07 &  0.94  &  0.005 &     0.004 &     0.44 &   0.40    \\ 
\tableline
FeII$_{opt}$     & r& 0.15  &    1  & {\bf -0.41}&   -0.22 &      0.21 &  -0.11  & 0.02 &     0.17  &     -0.004 & 0.04   \\ 
                  & P& 0.01  &  0  &   {\bf 2.1E-13} & 1.3E-4&     3E-4 &   0.07  & 0.77 &     0.003 &     0.95 &   0.48   \\ 
\tableline 
  $[$O III$]$ & r& -0.11  &  {\bf  -0.41 } &    1 &  {\bf  0.38 }&        0.30 & {\bf  0.56} & {\bf  0.44}&    {\bf 0.35  } &     {\bf  0.38 } &  {\bf 0.39 } \\ 
              & P& 0.06  &  {\bf  2.1E-13 }&   0 &  {\bf  1E-11} &    1.1E-7 & {\bf 0 } & {\bf  2.2E-15} &    {\bf 3.9E-10  }&   {\bf   1.9E-11}  & {\bf  3.5E-12 }  \\ 
\tableline
MgII$_{total}$    & r& {\bf 0.39 } &    -0.22  &  {\bf 0.38} &   1 &    -0.04 &   {\bf 0.50 } &   0.06 &     {\bf 0.42 } &     0.11 &   0.30    \\
                  & P& {\bf 3.2E-12 } &   1.3E-4  & {\bf  1E-11 }&  0 &        0.44 &   {\bf 0 } &        0.30 &    {\bf  7.5E-14} &     0.07 &   9.6E-8    \\

\tableline
H$\beta_{NLR}$ & r& -0.10  &    0.21 &    0.30 &   -0.04 &        1 &   0.09  & {\bf   0.70 }&     0.22  &   {\bf   0.56} &   0.28   \\ 
               & P& 0.07  &    3E-4  &  1.1E-7 &   0.44 &        0 &   0.14  & {\bf   0 }&     1.7E-4  &   {\bf   0 }&   7.4E-7     \\ 
\tableline
H$\beta_{broad}$ & r& 0.005 &    -0.11  & {\bf 0.56} &  {\bf 0.50} &        0.09 &   1  &   0.20 &  {\bf   0.77}  &     0.20& {\bf  0.55 }  \\ 
                 &P&  0.94  &    0.07  & {\bf 0} &  {\bf 0 }&      0.14 &   0 &   4.9E-4 &  {\bf    0 } &     6.8E-4 &  {\bf 0 }  \\ 
\tableline
H$\gamma_{NLR}$ & r& -0.16 &    0.02  & {\bf 0.44 }&   0.06 &      {\bf  0.70 }&  0.20  &  1 &     0.28  &   {\bf  0.61 }&  {\bf 0.36 }  \\ 
                & P& 0.005  &    0.77 &  {\bf 2.2E-15 }&   0.30 &      {\bf  0} &   4.9E-4  &  0 &     1.3E-6  &   {\bf  0 }& {\bf  3.2E-10 } \\ 
\tableline
H$\gamma_{broad}$ &r&  0.17  &    0.17  &  {\bf  0.35}&  {\bf 0.42 }&              0.22 & {\bf   0.77 } &   0.28 &     1  &     0.21 &   {\bf 0.65}    \\ 
                  &P&  0.004  &    0.003  & {\bf 3.9E-10} &  {\bf 7.5E-14}&  1.7E-4 &    {\bf 0} &    1.3E-6 &     0 &     2.1E-4 &   {\bf 0 }  \\ 
\tableline
H$\delta_{NLR}$ & r& -0.04 &    -0.004  & {\bf  0.38} &   0.11 &       {\bf   0.56 }&   0.20  & {\bf   0.61} &     0.21  &     1 &   0.22   \\ 
               & P& 0.44  &    0.95  & {\bf  1.9E-11} &   0.07 &      {\bf    0 }&   6.8E-4  & {\bf   0 }&     2.1E-4  &     0 &   1.7E-4    \\ 
\tableline
H$\delta_{broad}$ & r& -0.05 &    0.04  & {\bf 0.39 }&  0.30 &             0.28  &   {\bf  0.55} & {\bf  0.36 } & {\bf  0.65 }&     0.22  &     1     \\ 
                 & P& 0.40 &    0.48  &  {\bf 3.5E-12 } &9.6E-8 &       7.4E-7  &     0 &   {\bf 3.2E-10 }&  {\bf 0} &     1.7E-4 &     0    \\

\enddata

\tablenotetext{a}{Only multiplets 60 and 61 are included.}
\tablenotetext{b}{Optical Fe II is in range $\lambda\lambda$ 4000-5500 \AA}
\tablenotetext{c}{MgII$_{total}$ includes EWs of MgII core and MgII wings.}
\tablenotetext{d}{H$\beta_{broad}$, H$\gamma_{broad}$ and H$\delta_{broad}$ are summ of ILR and VBLR components for each line.}
\end{deluxetable}


\clearpage

\begin{deluxetable}{cccccccc}
\tabletypesize{\scriptsize }
\tablecaption{The correlations between EWs and the widths of the emission lines. The
Spearman coefficient of correlation (r) and P-value  are given in bold print for correlations with P$<$1E-9. \label{tbl-9}}

\tablehead{
\colhead{} & \colhead{} &
\colhead{EW Fe II$_{opt}$  } &\colhead{ EW Fe II$_{UV 60}$} &\colhead{EW H$\beta$ NLR } &\colhead{ EW H$\beta$ broad }&
\colhead{  EW [O III]} &\colhead{EW Mg II } 
}
 
\startdata

w NLR&r & 0.24& -0.02 &{\bf 0.52 }& -0.13& -0.01&-0.14\\
&P & 2.3E-5 &0.65 &{\bf 0 }& 0.03&0.86&0.02\\
\tableline
FWHM H$\beta_{broad}$ &r &{\bf -0.44}& -0.003 &-0.23&0.07& 0.09&0.14\\
&P &{\bf 4.9E-15} &0.96 &7.5E-5&0.2&0.12&0.01\\
\tableline
FWHM Mg II&r &{\bf -0.45}&0.13 &{\bf  -0.37}&0.1& 0.12& 0.32\\
&P &{\bf 8.9E-16}&0.03 &{\bf  4.0E-11}&0.1&0.04&1.5E-8\\
\tableline
w Fe II$_{opt}$&r & -0.33&0.01 &-0.34 &0.17&0.15&0.17\\
&P &7.9E-9& 0.83&3.2E-9&0.003&0.01&0.002 \\
\tableline
w Fe II$_{UV}$&r  &-0.20&{\bf 0.40} &-0.19&0.03&0.05&0.02 \\
&P &7.6E-4&{\bf 6.7E-13} &0.001&0.59&0.42&0.69\\

 \tableline
 
 \enddata
\end{deluxetable}


\clearpage

\begin{table}
\begin{center}
\tabletypesize{\tiny}
\caption{The correlations between line flux ratios and line parameters. The Spearman coefficient of correlation 
(r) and P-value  are given in bold print for correlations with P$<$1E-9. \label{tbl-8}}

\begin{tabular}{c c c c c c c c}
 \tableline\tableline

 &    & $\frac{\mathrm{Fe II}_{opt}}{\mathrm{Fe II}_{UV 60}}$& $\frac{\mathrm{Fe II}_{opt}}{\mathrm{Mg II}}$& $\frac{\mathrm{Fe II}_{UV}}{\mathrm{Mg II}}$&$\frac{\mathrm{Mg II}}{\mathrm{H\beta}_{broad}}$&$\frac{\mathrm{[O III]}}{\mathrm{H\beta}_{NLR}}$& $\frac{\mathrm{H\beta}_{broad}}{\mathrm{H\gamma}_{broad}}$\\
 \tableline\tableline
 \vspace{-4mm}
width NLR&r   &0.21&0.24 &0.12 &-0.036&{\bf -0.49 } & -0.16    \\
&P            & 2.2E-4 & 3.3E-5& 0.03 &0.54 &{\bf 0 } & 0.005  \\
\tableline
\vspace{-4mm}
FWHM H$\beta$ &r & {\bf -0.43}&{\bf -0.45}&-0.19 & 0.19&0.32&{\bf 0.44}\\
&P  & {\bf 2E-14 }& {\bf 4.4E-16}& 0.001&0.001&1.2E-8&{\bf 3.5E-15}\\
\tableline
\vspace{-4mm}
FWHM Mg II&r & {\bf -0.55}& {\bf -0.57}&-0.23 & {\bf 0.36}&{\bf 0.50}&{\bf 0.40}\\
&P & {\bf 0} &{\bf 0}&8.6E-5 & {\bf 1.75E-10}&{\bf 0}&{\bf 5.8E-13}\\
\tableline
\vspace{-4mm}
width Fe II$_{opt}$&r &-0.31& {\bf -0.37}&-0.17 &0.09&{\bf 0.43}&{\bf 0.48}\\
&P &8.5E-8&{\bf 1E-10}&0.004 &0.10&{\bf 1.2E-14}&{\bf 0}\\
\tableline
\vspace{-4mm}
width Fe II$_{UV}$&r  &{\bf -0.44} &-0.18&0.32  &0.065&0.23&0.14\\
&P & {\bf 1.5E-15}& 0.002&2.2E-8&0.26&5.2E-5&0.01\\
\tableline
\vspace{-4mm}
EW Fe II$_{opt}$&r & {\bf 0.69}& {\bf  0.76} &{\bf 0.39} &-0.13&{\bf -0.45}& {\bf -0.49}\\
&P &{\bf 0}&{\bf 0} &{\bf 4.3E-12 }&0.02&{\bf 8.9E-16}&{\bf 0}\\
\tableline
 \vspace{-4mm}
EW Fe II$_{UV 60}$&r & {\bf -0.42}& -0.07 &{\bf 0.52}  &{\bf 0.35}&0.02& -0.28\\
&P &{\bf 8.3E-14}& 0.26 &{\bf 0 } &{\bf 6.1E-10}&0.78&6.8E-7\\
\tableline 
\vspace{-4mm}
EW [O III]&r & -0.28&  {\bf -0.48}&{\bf -0.44}  &-0.09&{\bf 0.36 } & 0.34 \\
&P &6.9E-7& {\bf 0 }&{\bf  1.5E-15 }  &0.13& {\bf 2.0E-10 }&1.3E-9\\
\tableline
\vspace{-4mm}
EW H$\beta_{broad}$&r &-0.17&  {\bf -0.35}&{\bf -0.39} &-0.24&0.30 &{\bf 0.37}\\
&P &0.003&{\bf 4.5E-10} &{\bf 2.0E-12} &2.4E-5&1.6E-7&{\bf 3.2E-11}\\
\tableline
\vspace{-4mm}
EW Mg II$_{total}$ &r & {\bf -0.50}&  {\bf -0.70}&{\bf -0.51} &{\bf 0.57}&0.30& 0.19\\
&P &{\bf 0}&{\bf 0}&{\bf 0} &{\bf 0}&1.0E-7&0.001\\
\tableline
\vspace{-4mm}
$\frac{\mathrm{Fe II}_{opt}}{\mathrm{Fe II}_{UV 60}}$ &r & 1&  {\bf 0.82}&0.06 &{\bf -0.45}&{\bf -0.40}& -0.26\\
&P &0 &{\bf 0}&0.29 &{\bf 6.7E-16}&{\bf 1.3E-12}& 7.3E-6 \\
\tableline
\vspace{-4mm}
$\frac{\mathrm{Fe II}_{opt}}{\mathrm{Mg II}}$ &r & {\bf 0.82}& 1&{\bf 0.57}  &{\bf -0.50}&{\bf -0.47}& {\bf -0.46} \\
&P &{\bf 0}& 0 &{\bf 0} &{\bf 0}&{\bf 0}&{\bf 0}\\
\tableline
\vspace{-4mm}
$\frac{\mathrm{Fe II}_{UV}}{\mathrm{Mg II}}$ &r & 0.06& {\bf 0.57}& 1 &-0.20&-0.26& {\bf -0.43} \\
&P &0.29& {\bf 0 }&0 &5.5E-4&5.0E-6&{\bf 6.9E-15}\\
\tableline
\vspace{-4mm}
$\frac{\mathrm{Mg II}}{\mathrm{H\beta}_{broad}}$ &r & {\bf -0.45}&  {\bf -0.50}&-0.20&1&0.07& -0.16 \\
&P &{\bf 6.7E-16}&{\bf 0}&5.5E-4&0&0.23 &0.006\\
\tableline
\vspace{-4mm}
$\frac{\mathrm{[O III]}}{\mathrm{H\beta}_{NLR}}$ &r & {\bf -0.40}&  {\bf -0.47}&-0.26&0.07&1& {\bf 0.37} \\
&P                              &{\bf 1.3E-12}&{\bf 0}    &5.0E-6&0.23&0&{\bf 4.9E-11} \\
\tableline
\vspace{-4mm}
$\frac{\mathrm{H\beta}_{broad}}{\mathrm{H\gamma}_{broad}}$ &r &-0.26&  {\bf -0.46}&{\bf -0.43 }  &-0.16&{\bf 0.37 } & 1\\
&P &7.3E-6&{\bf 0}&{\bf  6.9E-15 }&0.006&{\bf 4.9E-11}&0 \\
\tableline

\end{tabular}
\end{center}
\end{table}

\clearpage
\begin{deluxetable}{ccccccc}
\tablecaption{The average values and standard deviations (SD) for the line flux ratios, calculated for the total sample 
(293 objects), and for subsamples
with $\mathrm{log([O III]/H\beta}_{NLR}\mathrm{)>0.5}$ (AGN dominant, 247 objects ) and $\mathrm{log([O III]/H\beta}_{NLR}\mathrm{)<0.5}$ (starburst - 
SB dominant, 46 objects). The ratio $\mathrm{O III}/\mathrm{H\beta_{NLR}}$
is shown in Table in logaritmic form because of the large dispersion in the case of H$\beta_{NLR}\approx$0.
 \label{tbl-nova2}}
\tablehead{

\colhead{} & \multicolumn{2}{c} { ALL}  &  \multicolumn{2}{c}{AGN dominant}&  \multicolumn{2}{c}{SB dominant} \\
\colhead{} & \colhead{average values} & \colhead{ SD} & \colhead{average values} & \colhead{ SD} & \colhead{average values} & \colhead{ SD} 
 }
\startdata

 $\frac{\mathrm{Fe II}_{opt}}{\mathrm{Fe II}_{UV 60}}$ &  5.482    & 2.881  &    5.114  & 2.776 & 7.455  & 2.650   \\    
     \tableline
$\frac{\mathrm{Fe II}_{opt}}{\mathrm{Mg II}}$ &    0.902     &  0.524 &     0.804 &  0.472&  1.428&  0.476  \\
\tableline
 $\frac{\mathrm{Fe II}_{UV 60}}{\mathrm{Mg II}}$   &  0.166    & 0.062 &    0.160  & 0.060 & 0.200 & 0.060    \\ 
  \tableline
  $\frac{\mathrm{Mg II}}{\mathrm{H\beta}_{broad}}$    &     1.938     & 1.048  &    1.959   & 1.108& 1.827 & 0.627    \\ 
         \tableline 
 $\frac{\mathrm{H\beta}_{broad}}{\mathrm{H\gamma}_{broad}}$  &     2.042    & 0.345 &    2.109  & 0.287& 1.683 & 0.409     \\  
       \tableline
$\mathrm{log}\frac{\mathrm{[O III]}}{\mathrm{H\beta}_{NLR}}$     &     0.979    &  0.769 &     1.148 &  0.695 &  0.071&  0.442   \\

\tableline

 \enddata
\end{deluxetable}

\begin{deluxetable}{ccccc}
\tabletypesize{\scriptsize }
\tablecaption{The percentage of the objects from the sample with better/worse fit with new iron templates (based on H$\beta$ or Mg II profiles) comparing the initial iron template. 
The quality of fit is measured by differences between the $\chi^2$ from the fits ( $\chi_{old}^2$ - the initial (single Gaussian) iron template, $\chi_{new}^2$ - the iron templates with H$\beta$ or Mg II profile). \label{App}}

\tablehead{
\colhead{ } & \multicolumn{4}{c}{\bf Percentage of the objects from the sample }\\
\colhead{} &\colhead{ Fe II$_{opt}$  (H$\beta$ profile)} &\colhead{Fe II$_{opt}$   (Mg II profile) } &\colhead{Fe II$_{UV}$ (H$\beta$ profile)  }&\colhead{ Fe II$_{UV}$ (Mg II profile)  }  }

\startdata

 $\chi_{new}^2 > \chi_{old}^2$ & 15.17 \% & 50.00 \% & 82.59 \% & 89.73 \% \\
 (worse fit)& &  &  &  \\
 \tableline
 $\chi_{new}^2 < \chi_{old}^2$ & 84.83 \% & 50.00 \% & 17.41 \% & 10.27 \% \\
 (better fit)& &  &  &  \\
 \tableline

 \tableline
 
 \enddata
\end{deluxetable}

   \end{document}